\documentclass[trackchanges]{aastex701}

\usepackage{amsmath,amssymb,amsfonts}%

\begin{document}

\title[Black hole Limits Redefined]{Black hole Limits Redefined: Extreme Efficiency in Black Hole Jets}

\author[orcid=0000-0002-1655-9912,sname='Antonios']{Antonios Nathanail}
\affiliation{Research Center for Astronomy and Applied Mathematics, Academy of Athens, Soranou Efesiou 4, Athens, 11527, Greece}
\email[show]{anathanail@Academyofathens.gr}  

?

\begin{abstract}

Relativistic jets from black holes can extract energy not only from accretion but also directly from the black hole’s spin, as described by the Blandford–Znajek mechanism. A longstanding question is whether magnetic flux can accumulate near the event horizon to such an extent that it halts accretion entirely, enabling energy extraction purely from spin. Previous studies have shown that accretion persists through instabilities and that jet power only modestly exceeds the accretion energy budget, yet some observational results suggest much higher efficiencies. Here we present state-of-the-art general relativistic magnetohydrodynamic (GRMHD) simulations that sustain a quasi-steady magnetically arrested disk state for approximately 10,000 dynamical times, during which accretion is globally suppressed across the full azimuthal extent. In this regime, jet power exceeds the accretion energy input by more than two orders of magnitude, demonstrating a previously unachieved level of efficiency. These results challenge conventional assumptions about the limits of black hole energy extraction and suggest a new framework for interpreting powerful jet systems. Our findings raise important questions about the long-term stability of such states and the fundamental limits of the Blandford–Znajek process.
\end{abstract}





\section{Introduction}\label{sec1}

Relativistic jets originating from black holes are among the most energetic 
phenomena in the universe. These jets are typically thought to extract energy 
primarily from the accretion process \citep{Blandford1974,  Narayan2008,  Igumenshchev2008}, where infalling matter powers the creation 
(or advection) of magnetic fields that launch material away from the black hole.
The accumulation of magnetic flux by the accreting material enables the field to exert a torque on the spinning black hole, facilitating the extraction of rotational energy. In the standard equipartition framework, the magnetic field strength squared is expected to scale with the accretion rate, implying a direct correlation between jet power and accretion luminosity. Although such a correlation is supported by many observations \citep{Rawlings1991, Celotti1993, Punsly2006, Celotti1997}, there are clear cases where the jet power vastly exceeds the accretion luminosity \citep{Maraschi2003, Ghisellini2010, Ghisellini2014, Chen2023}. This suggests the operation of a highly efficient extraction process, likely driven by black hole spin alone rather than accretion.
The Blandford–Znajek mechanism \citep{Blandford1977} suggests that 
black hole spin itself can also be a significant energy source for jet 
production. Previous studies \citep{Narayan2003, Tchekhovskoy2011}
have demonstrated that jet power can slightly exceed the available energy from 
accretion, but recent observations suggest that jet efficiencies can greatly 
surpass theoretical expectations  \citep{Ghisellini2014, Ghisellini2017, 
Chen2023jets, Li2024, He2024}. In 
some cases, the jet power is observed to exceed the available accretion energy 
by orders of magnitude.

Observational studies have found that in radio-loud active galactic nuclei, and particularly in Flat Spectrum Radio Quasars, jet power can 
often exceed the accretion luminosity. \citet{Ghisellini2014} showed that in 
some powerful jets, $P_{\rm Jet}/L_{\rm disk}$, where $P_{\rm Jet}$ 
is the power of the jet and $L_{\rm disk}$ is 
the accretion disk luminosity, can reach values greater than 10, with some cases 
going to 100 and beyond. This suggests that jets in these systems are highly 
efficient in extracting rotational energy from the black hole, often facilitated 
by strong magnetic fields in Magnetically Arrested Disks  \citep{Zamaninasab2014}. 

An accretion disk, forming around black holes across all mass scales, provides a 
reservoir of matter that inherently carries magnetic fields, either advected 
with the inflowing gas or generated in situ through battery processes 
\citep{Contopoulos1998, Contopoulos2015, Contopoulos2018}. While the magnetic field 
within the disk itself is typically weak and non-dynamical—characterized by a 
magnetization parameter $\sigma = B^2/\rho c^2 < 10^{-2}$ —accretion 
leads to the gradual buildup 
of a magnetized region near the black hole. Equipartition is reached when the 
magnetic pressure exerted by the accumulated flux on the event horizon balances 
the ram pressure of the infalling gas, forming a magnetic barrier, or "funnel," 
that can, in principle, suppress accretion. At this point the state is described 
as a magnetically arrested disk, and the normalized magnetic flux on the 
event horizon ($\phi_{\rm BH}= \Phi_{\rm BH}/\sqrt{\dot{m}}\approx 50$, 
see Appendices for formal 
definitions) saturates and accretion cannot bring more magnetic field onto the 
black hole \citep{Tchekhovskoy2011}. However, early numerical simulations 
\citep{Igumenshchev2008} demonstrated that accretion persists through non-
axisymmetric instabilities, commonly identified as interchange instabilities 
(including the Kruskal-Schwarzschild and/or magnetized Rayleigh-Taylor modes).

Thus, the fundamental question we address in this study is whether a steady-state (or quasi-steady) solution exists in which accretion is entirely halted across the event horizon. The answer to this question is tightly connected with the limits of the efficiency of energy extraction from a spinning black hole. The upper limit of jet efficiency remains uncertain, where jet efficiency is defined as the ratio of the energy outflowing in the jet to the rest-mass energy input via accretion, $\eta =P_{\rm Jet}/\dot{m} c^2$. An established feature of the magnetically arrested disk state is that the efficiency of the generated jet can surpass the accretion energy, 
$\eta \approx 140\%$, 
in average for thousands of $t_{\rm g}$ \citep{Tchekhovskoy2011}. Furthermore, a connection has been established between the efficiency and the jet structure with the spin a black hole \citep{Narayan2012, Narayan2022}. Despite advancements in simulation techniques, no study has yet demonstrated a stable, quasi-steady magnetically arrested disk state over long timescales in which accretion is globally suppressed, leaving the regime of fully halted accretion, largely unexplored. Here, we present results from high-resolution simulations that maintain a quasi-steady magnetically arrested disk state for approximately 10,000 dynamical times till the end of the simulation time, during which accretion is fully suppressed over a full azimuthal extent. We show that in this regime, jet power can exceed the energy input from accretion by more than two orders of magnitude, thus providing an unprecedented level of efficiency.

\section{Results}\label{sec2}

We perform high-resolution GRMHD simulations of black hole accretion using the BHAC code \citep{Porth2017, Olivares2020}, focusing on the formation and evolution of magnetically arrested disks (MADs) around rapidly spinning black holes ($a_* = 0.9375$). The simulations begin with a weakly magnetized, radially infalling gas configuration. As accretion proceeds, the buildup of magnetic flux near the horizon leads to the MAD state and the launching of relativistic jets.

We consider four models that differ in initial magnetic pressure relative to gas pressure, characterized by initial $2p_{\rm max}/B^2_{\rm max}$ values of 100, 26, 13, and 7. The corresponding maximum magnetization in the torus spans $\sigma_{\rm max} = 2 \times 10^{-4}$ to $1.6 \times 10^{-3}$, changing by a factor of 2 for each model, labeled MAD.S.100, MAD.S.26, MAD.S.13, and MAD.S.7. Details on the numerical setup are provided in the Methods.



\begin{figure}[h]
\centering
\includegraphics[width=0.485\textwidth]{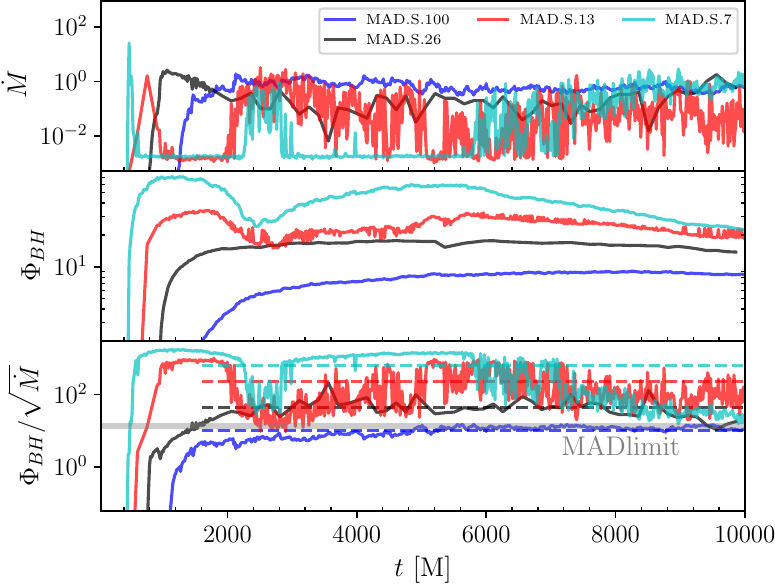}
\includegraphics[width=0.47\textwidth]{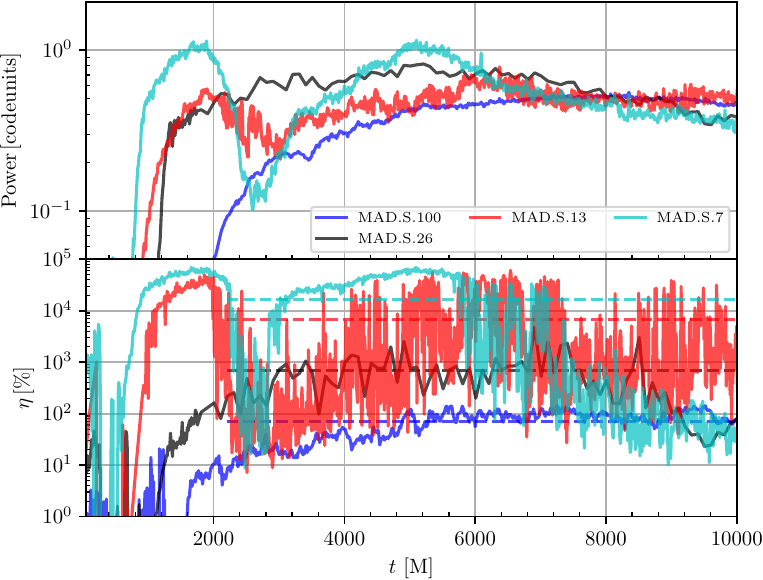}
\caption{Left Panels: Upper panel: evolution of the mass accretion measured across the black hole event horizon. Middle panel: evolution of the magnetic flux accreted onto the black hole. Lower panel: evolution of the normalized magnetic flux accumulated on the black-hole horizon. Right panels: Upper panel: the power of the jet. Lower panel: the efficiency of the jet for each magnetically arrested disk model. The dashed lines correspond to the average of the quantity from 1,500 $t_{\rm g}$ (2,000 $t_{\rm g}$ for the right one) till 10,000  $t_{\rm g}$.}
\label{mdot-p}
\end{figure}

We monitor the magnetic field strength, the mass accretion rate, and the total energy flux in the jet to assess the system's efficiency. 
Additional validation tests on floor treatment and inversion methods, are detailed in the Methods section.
Fig. \ref{mdot-p} shows the temporal evolution of the rest-mass accretion rate $\dot{M}$, accumulated magnetic flux on the event horizon $\Phi_{\rm BH}$, and normalized magnetic flux $\phi_{\rm BH}=\Phi_{\rm BH}/\sqrt{\dot{m}}$ for four models: a standard magnetically arrested disk model and three increasingly extreme magnetically arrested disk configurations. 
In the standard magnetically arrested disk model (MAD.s.100), $\dot{M}$ fluctuates but remains non constant and nonzero, meaning that accretion never stops, consistent with prior studies. However, in the most extreme model (MAD.s.7), $\dot{M}$ drops to a very low and constant value between 3,000 and 6,000 dynamical times ($t_{\rm g}$). Simultaneously, $\phi_{\rm BH}$ reaches unprecedented levels, exceeding 50 times the magnetically arrested disk (MAD) limit reported in the literature. Animations for models MAD.s.7 and MAD.s.13 can be found \href{https://zenodo.org/records/15745329?token=eyJhbGciOiJIUzUxMiJ9.eyJpZCI6IjI2YzQ5NDg3LTRmM2EtNGQ1NC04NGVmLTkxMzc2MzM5YWMyNCIsImRhdGEiOnt9LCJyYW5kb20iOiI4YzlkZjJlMmY4YTAzMWUwY2YzYjMzZWFlMzY3NGM4NSJ9.fNqJn0kodbH1q7m6FpHITICHW4mOsZkPcLNcTPiIydaH76QNPMjXKchFsWEfhmP6AAL5YaAYXJERbye2muRl-w}{here} and \href{https://zenodo.org/records/15745251?token=eyJhbGciOiJIUzUxMiJ9.eyJpZCI6Ijg4ZTc2ZWRhLWU2MzItNGU3Ny1hZTcwLTViYmI4ODBjZWNjNSIsImRhdGEiOnt9LCJyYW5kb20iOiJhMjZlMGY5ZTI0MDQyOGNjZTE1NzcxMjQxMWNhZTBlZCJ9.K5D5dAZdwx99JE0SKBVAsbq4p0rZ-KZFYw--D4to9X88iLvcSkpyyUBU5p0B7z30n31dRUIkXX4Sa1qxOMG5TQ}{here} respectively.

\begin{figure}[h]
\centering
\includegraphics[width=0.49\textwidth]{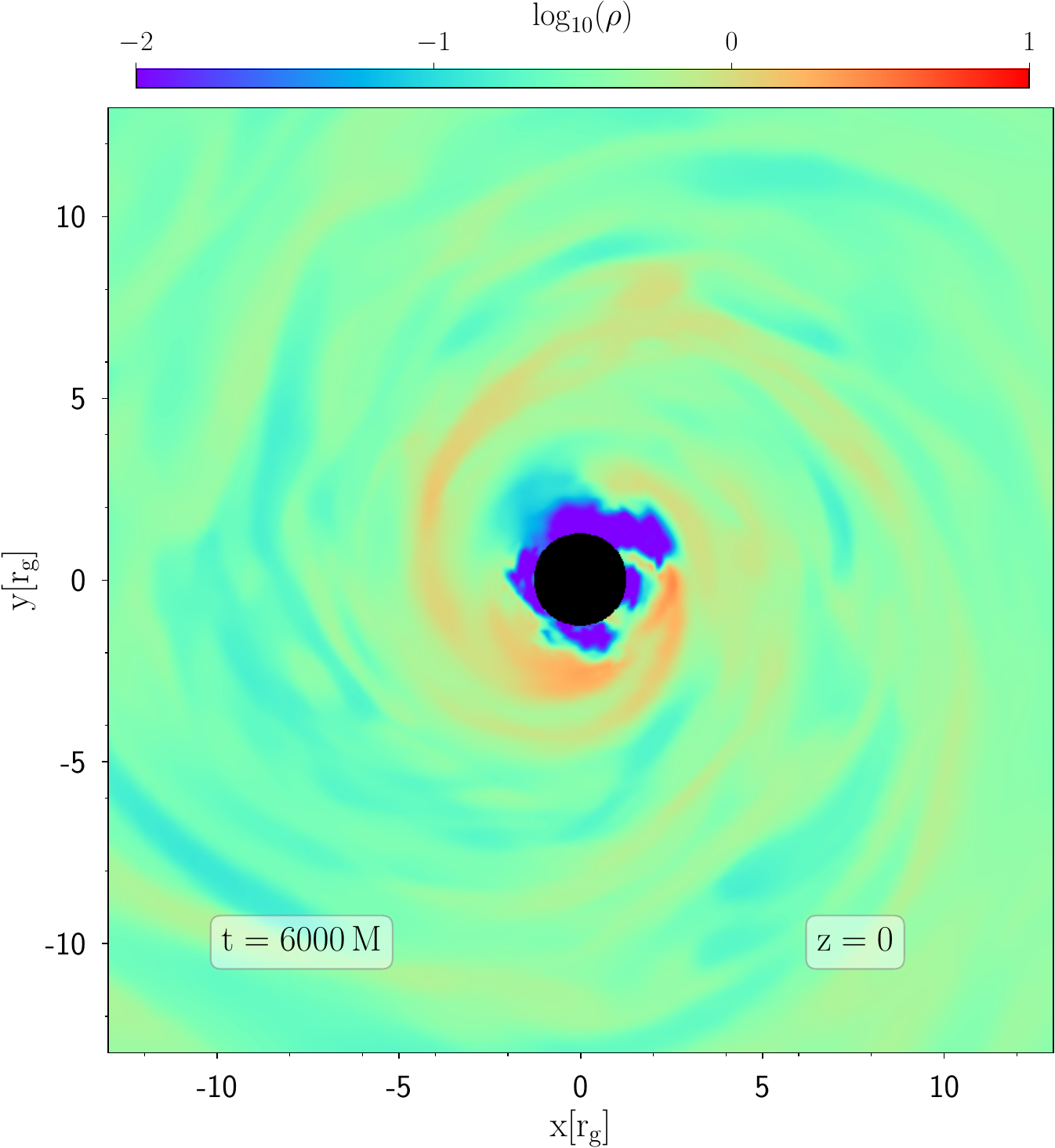}
\includegraphics[width=0.49\textwidth]{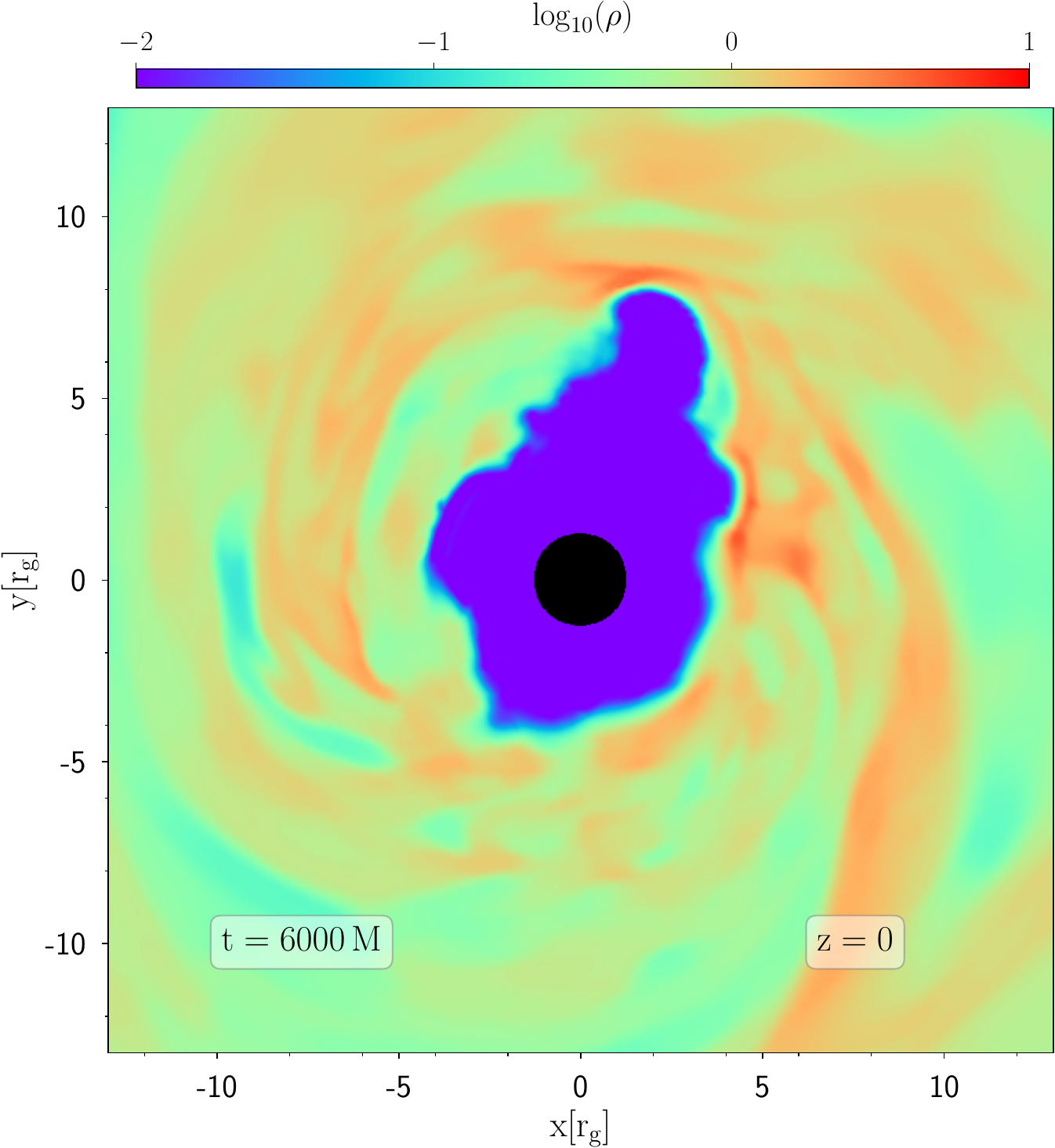}
\caption{A 2D slice of the simulation at the equatorial plane showing rest-mass density  at 6,000  $t_{\rm g}$. Left panel: model MAD.S.36 and Right panel: the extreme model MAD.S.7. The accretion disk in the extreme model has been expelled continuously for more than 3,000 dynamical times.}
\label{equa}
\end{figure}

Fig. \ref{equa} presents a 2D slice of the simulations at the equatorial plane for model MAD.s.26 and the extreme model MAD.S.7. In MAD.S.26 (left panel), accretion is highly disrupted, with inflowing matter being pushed back at several azimuthal locations. However, intermittent streams of infalling gas still manage to penetrate the magnetic barrier and reach the event horizon. This behavior illustrates how non-axisymmetric instabilities allow accretion to persist, even in a strongly magnetized state.
In contrast, the extreme MAD.S.7 model (right panel) exhibits a strikingly different configuration. Matter is effectively halted several $r_{\rm g}$ from the black hole, forming a sustained accumulation region where the magnetic pressure prevents further inward motion. This state remains stable for approximately 3,000 $t_{\rm g}$, during which no significant mass inflow occurs. 

\begin{figure}[h]
\centering
\includegraphics[width=0.49\textwidth]{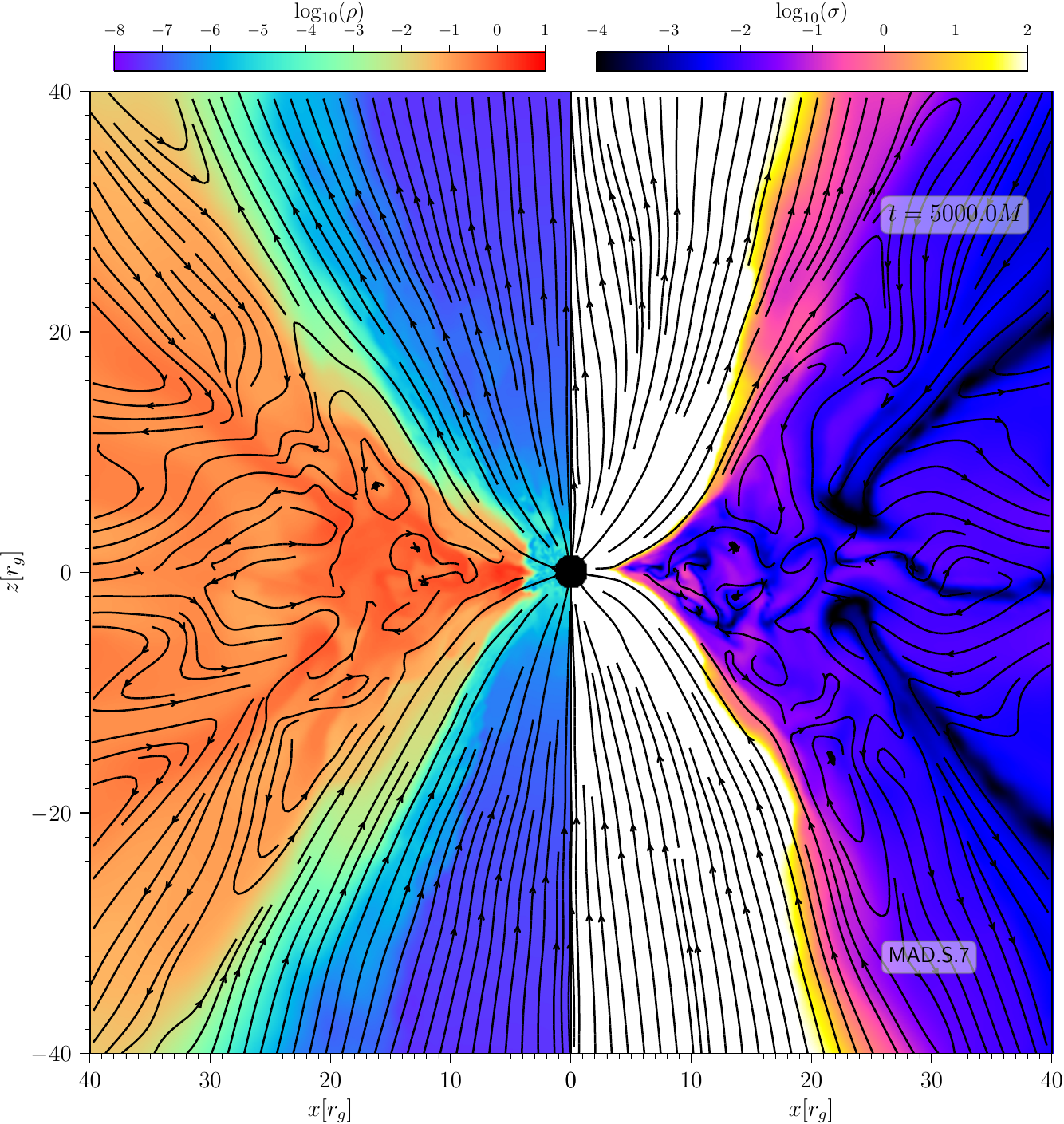}
\caption{A 2D slice of the simulation at constant azimuthal $\phi=180^o$ (left panel) and 
$\phi=0^o$ (right panel) for the extreme model MAD.S.7 at 5,000  $t_{\rm g}$. Left panel depicts the rest mass density whereas the right panel the magnetization $\sigma$.}
\label{vert}
\end{figure}

A vertical slice of constant azimuthal angle ($\phi = 0^o$)   is presented in Fig. 
\ref{vert}, illustrating the rest-mass density (left panel) and magnetization ($\sigma$, 
right panel) for model MAD.S.7. The figure highlights how the disk matter is regulated by 
the accumulated magnetic flux, effectively preventing accretion and keeping material 
stalled at a distance of approximately 5$r_{\rm g}$ from the event horizon. This extended 
standoff distance further reinforces the notion that the system has reached an extreme 
magnetically arrested disk state, where the magnetic barrier remains dominant over the ram 
pressure of the inflowing gas.
Notably, flux eruption events, which are typically observed in magnetically arrested disk 
systems \citep{Ripperda2022}, do not appear in our simulations in the usual form
(for a detailed discussion of flux eruption events in these super-MAD simulations, see the Methods section).

The efficiency with which a black hole extracts and converts energy is central to jet formation, shaping jet power across a range of environments. In standard MAD states, the Blandford–Znajek mechanism allows efficiencies up to 140\% of the accreted rest-mass energy. However, when accretion stalls and magnetic flux remains high, efficiencies can reach much more extreme values. As shown in the right panels of Fig.\ref{mdot-p}, all super-MAD configurations exhibit exceptionally high efficiencies, reaching ~6,000\% in model MAD.S.13. In the most extreme case, MAD.S.7, the efficiency exceeds 40,000\% during the period of suppressed accretion, indicating energy extraction at 400 times the rest-mass inflow rate.

\begin{figure}[h]
\centering
\includegraphics[width=0.485\textwidth]{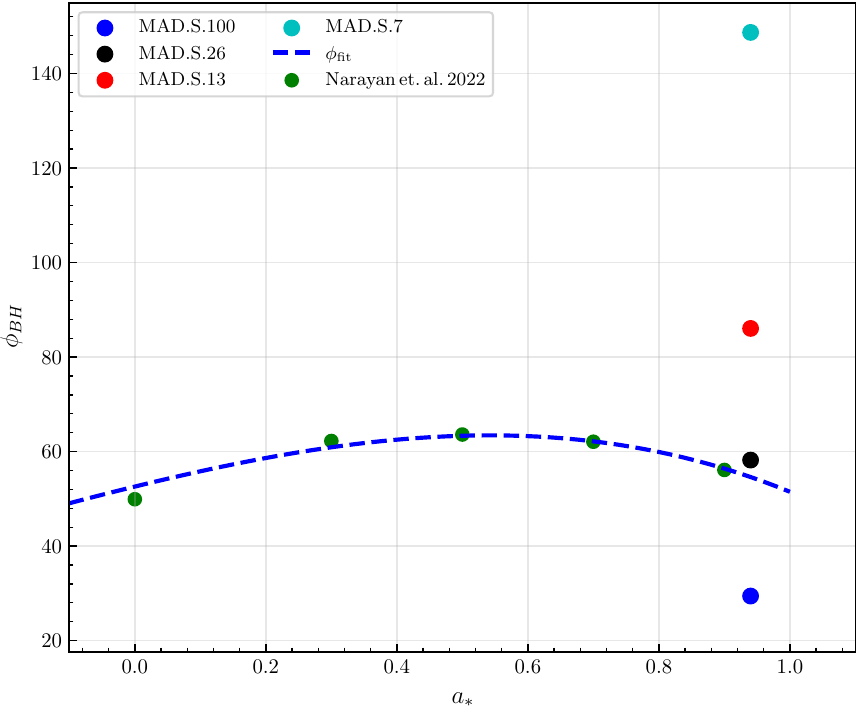}
\includegraphics[width=0.47\textwidth]{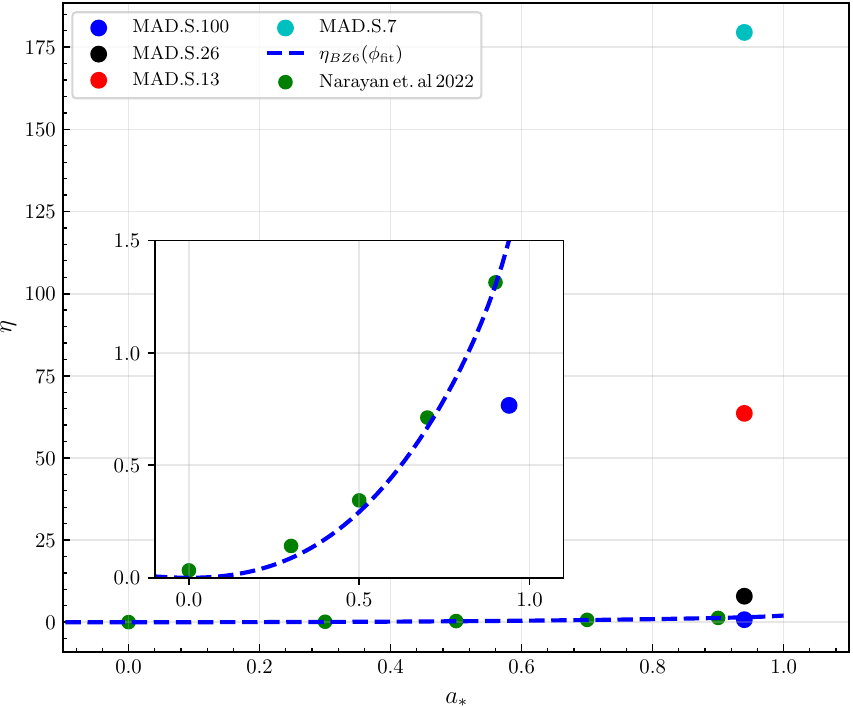}
\caption{
Left panel: Black hole spin versus the normalized 
magnetic flux at the event horizon. To match the 
results of \citet{Narayan2022} (green dots), the 
flux values from our simulations have been 
rescaled by a factor of $\sqrt{4\pi}$ (see Methods for details). 
The dashed blue line shows the fit from the green dots.
Right panel: Black hole spin versus the efficiency from our simulations and that of 
\citep{Narayan2022} in green. 
The dashed blue line represents the BZ6 model prediction for jet power (equation 10 in \citealt{Tchekhovskoy2010}), calculated by inserting the magnetic flux fitting function $\phi_{\rm fit}(a_*)$, which is taken from the left hand panel into the theoretical expression.
}
\label{spin-vs}
\end{figure}

Having established the extreme efficiencies observed in our simulations, it is now essential to place these results in the broader context of standard magnetically arrested disk models. Previous studies have demonstrated that magnetically arrested disk systems typically exhibit jet efficiencies around $\eta\approx$1.3–1.4 (\citet{Tchekhovskoy2011, McKinney2012} consider a model that has $\eta\approx 3$), with a well-defined correlation between black hole spin and magnetic flux saturation. However, our findings suggest that under certain conditions, 
black holes in extreme magnetically arrested disk states can reach dramatically higher efficiencies, challenging conventional expectations. Figure \ref{spin-vs}
presents the relationship between normalized magnetic flux $\phi_{\rm BH}$ and black hole spin (left panel), alongside jet efficiency $\eta$ as a function of spin (right panel). In the left panel, we compare our results with data from 
\citet{Narayan2022} and their previously derived fit.

The right panel shows the jet efficiency $\eta$ versus spin, again compared with \citet{Narayan2022} and the standard  formula prediction from \citet{Tchekhovskoy2010} ((equation 10 in \citealt{Tchekhovskoy2010}), calculated by inserting the magnetic flux fitting function $\phi_{\rm fit}(a_*)$, which is taken from the left hand panel into the theoretical expression). While standard magnetically arrested disk states typically achieve $\eta \approx 1.3-1.4$ \citep{Tchekhovskoy2011}, our extreme magnetically arrested disk models far exceed this value. The first extreme magnetically arrested disk configuration (MAD.S.26) reaches an average efficiency of $\eta\approx 8$, nearly six times higher than standard MADs. More strikingly, for the most extreme model (MAD.S.7), the efficiency surpasses $\eta>170$, implying that the system extracts energy on average at a rate 170 times the accreted rest-mass energy, an unprecedented scenario. To emphasize this contrast, the right panel includes an inline zoom-in plot, illustrating how standard magnetically arrested disks cluster around $\eta\approx 1.4$, while our extreme configurations reside at significantly higher efficiencies. These results further reinforce the notion that high-spin black holes in extreme magnetically arrested disk states can operate in a vastly more efficient energy extraction regime than previously thought.

\section{Conclusion}\label{sec13}

Our study reveals that extreme magnetically arrested disk states can 
reach unprecedented levels of magnetic flux accumulation and jet 
efficiency, far exceeding what has been previously observed in numerical 
simulations. Accretion can be suppressed for long time windows (of 
10,000 $t_{\rm g}$ for models MAD.s.26 and MAD.s.13 and 
3,000 $t_{\rm g}$ for MAD.s.7), 
coupled with extraordinary energy extraction from the black hole’s spin, 
suggests the possibility of a new, highly efficient mode of jet 
launching. However, whether these extreme  states represent long-
lived quasi-steady configurations or transient phases remains an open 
question.
If such extreme magnetically arrested states are realizable in astrophysical environments, their implications could be profound.
We note that the evacuated, force-free magnetospheric configuration that emerges in our
simulations closely resembles the theoretical picture recently outlined by \cite{Blandford2022}, 
who argued that highly magnetized black hole magnetospheres may evolve
toward states in which accretion becomes dynamically subdominant and spin–extraction
processes dominate. Our results provide a concrete numerical realization of this regime.
These results further suggest that some radio-loud active galactic nuclei may
operate in a regime where the jet power far surpasses the standard 
Blandford-Znajek efficiency. In X-ray binaries, transient episodes of 
extreme magnetically arrested disk states may account for the sudden 
flaring activity observed in some systems. Additionally, the extreme 
energy extraction seen in our simulations may offer a potential 
explanation for the ultra-powerful relativistic jets observed in gamma-
ray bursts, where accreted stellar material could naturally lead to an extreme MAD state.
Furthermore, our findings raise the question of how long black holes can sustain these energy extraction phases.  Future work should focus on long-duration simulations to determine whether such states can persist on astrophysical timescales.

\begin{acknowledgments}
The author were supported by the Hellenic Foundation for Research and Innovation (ELIDEK) under Grant No 23698, and by computational time granted from the National Infrastructures for Research and Technology S.A. (GRNET S.A.) in the National HPC facility - ARIS - under project ID 16033. 

\end{acknowledgments}

\software{BHAC \citep{Porth2017, Olivares2019},  
          }

\appendix

\section{Detailed Numerical Setup}

The initial conditions are set to ensure that the the 
inflowing material can build a magnetized funnel above the black hole and a relativistic 
jet can be launched extracting some of the black hole's spin energy. The magnetohydrodynamic equations 
governing the system include terms for energy and momentum conservation, with particular 
emphasis on the dynamics of the magnetic field and its interaction with the plasma. 
The initial numerical configuration consists of  a rotating  Kerr black hole  and  a
perturbed torus infused with a poloidal magnetic field. 
All simulations that we  conducted in this study are performed in three spatial dimensions, employing the general relativistic magnetohydrodynamic code BHAC (open source \url{https://bhac.science/}  \citet{Porth2017}). 
This code  exploits second-order shock-capturing finite-volume methods and has been  extensively tested \citep{Porth2017} and used in various studies \citep{Nathanail2018c,  Mizuno2018, Nathanail2020b}. It employs the constrained-transport method  \citep{DelZanna2007} to ensure a divergence-free magnetic field \citep{Olivares2019} and has undergone  testing and comparisons with similar-capability magnetohydrodynamic codes in general relativity \citep{Porth2019}. We explore  magnetically arrested disk configurations where we vary the initial magnetization of the torus. The matter around the black hole initially,  consist of an equilibrium torus with a constant  specific angular momentum of $\ell=6.76$  \citep{Fishbone76} orbiting around  a Kerr black hole with dimensionless spins of $a = 0.937$.

We initialize the magnetic field inside the torus with a nested loop that traces the contours of density, as in typical  SANE simulations \citep{Porth2019, Wong2021, Narayan2022}, 
and 
 follow the  simple form $A_{\phi}=\max \left(\frac{\rho}{\rho_{\max}} -0.2, 0\right)$.

 Note that the typical initialization for a magnetically arrested disk state \citep{Tchekhovskoy2011, Cruz2021, Ripperda2022, Nathanail2025} is  much different, and it has been shown that it does not evolve to configurations similar to the one described here  \citep{Cho2023, Cho2025}.

The net magnetic flux in the torus is zero however, the change of 
the initial magnetic field has the impact to increase the magnetic flux at the inner part 
of the torus from where accretion begins.

The computational domain uses  spherical logarithmic Kerr-Schild coordinates. 
The adiabatic index for all models is $\hat{\gamma}=4/3.$ 
In Table \ref{table:models} we report all simulations conducted in this study in 3D. The initial magentic field strength is set by the quantity  $2p_{\rm  max}/(B^2)_{\rm max}$, where the location of maximum of fluid pressure and magnetic pressure may not coincide. 
\begin{figure}[h]
\centering
\includegraphics[width=0.76\textwidth]{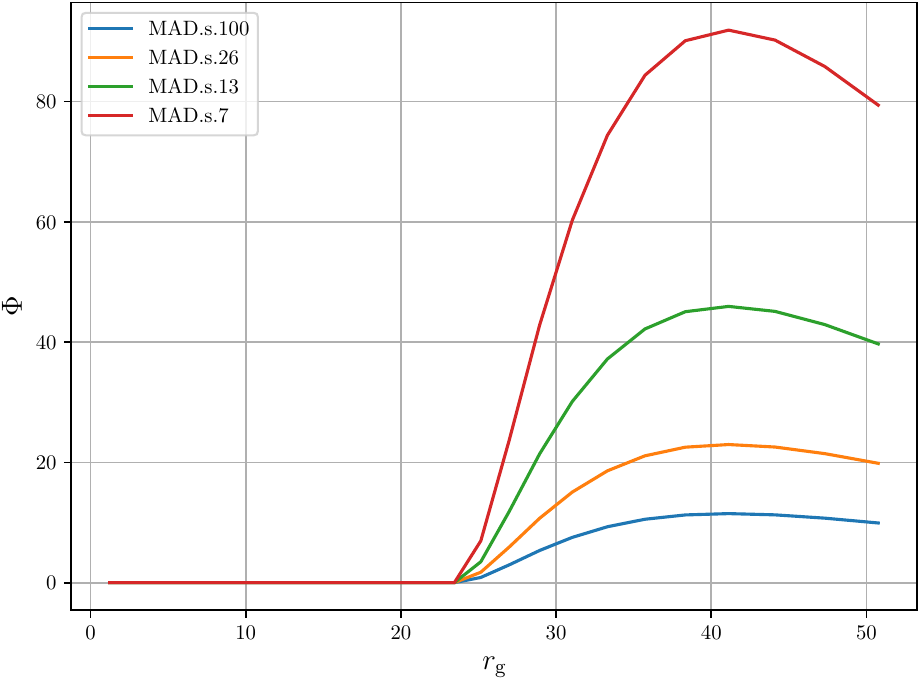}
\caption{Magnetic flux $\Phi$ integrated across r in the initial time.}
\label{initB}
\end{figure}

\begin{table*}
  \centering
   \caption{Initial parameters for the models considered. The first
        column is the name of the model, whereas the second column
         the initial maximum values of $2p/B^2$ and the third column  the initial maximum magnetization $\sigma_{\rm max}$. 
        The next column reports the resolution of each run, whereas columns 5 and 6  report the inner 
radius and maximum density radii of the torus, and the 7th column the outer
radius of the domain, in units of $r_{\rm g}$. The  last column report the average MRI quality
factor $Q_{\theta}$ at the heart of the torus \citep{Porth2019_etal}.
}
        \begin{tabular}{|l|c|c|l|c|c|c|c|} \hline \hline
                model  &  \underline{$2p_{\rm max}$}  &
                $\sigma_{\rm max}$ & $N_r \times N_{\theta} \times N_{\phi}$
                & $r_{\rm in}$ & $r_{\rm p_{\rm max}}$ & $r_{\rm max}$
                &$\langle Q_{\theta}\rangle$ \\
                &$(B^2)_{\rm max}$&$\times10^{-4}$&&&&& \\
                \hline
        MAD.S.100 & $100$   &  $2$ & $384 \times 192 \times 192$ 
        &20&41&2500&$>10$\\
        MAD.S.26  & $26$    &  $4$   & $384 \times 192 \times 192$
        &20&41&2500&$>10$\\
        MAD.S.13   & $13$    &  $8$   & $384 \times 192 \times 192$
        &20&41&2500&$>10$\\
        MAD.S.7   & $7$    &  $16$   &  $384 \times 192 \times 192$
        &20&41&2500&$>10$\\
        
                \hline \hline
        \end{tabular} \label{table:models} \end{table*}

\begin{figure}[h]
\centering
\includegraphics[width=0.49\textwidth]{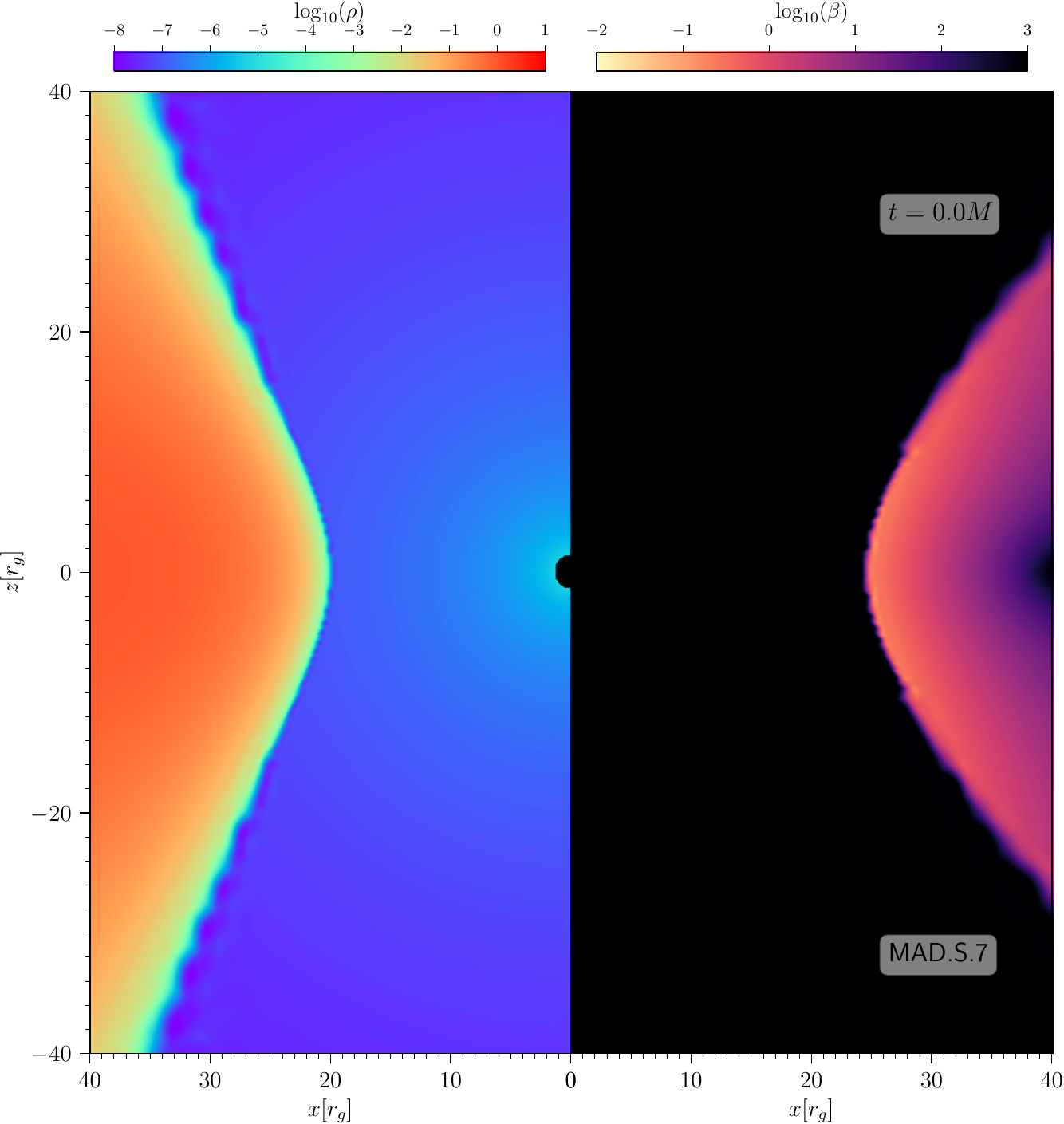}
\includegraphics[width=0.49\textwidth]{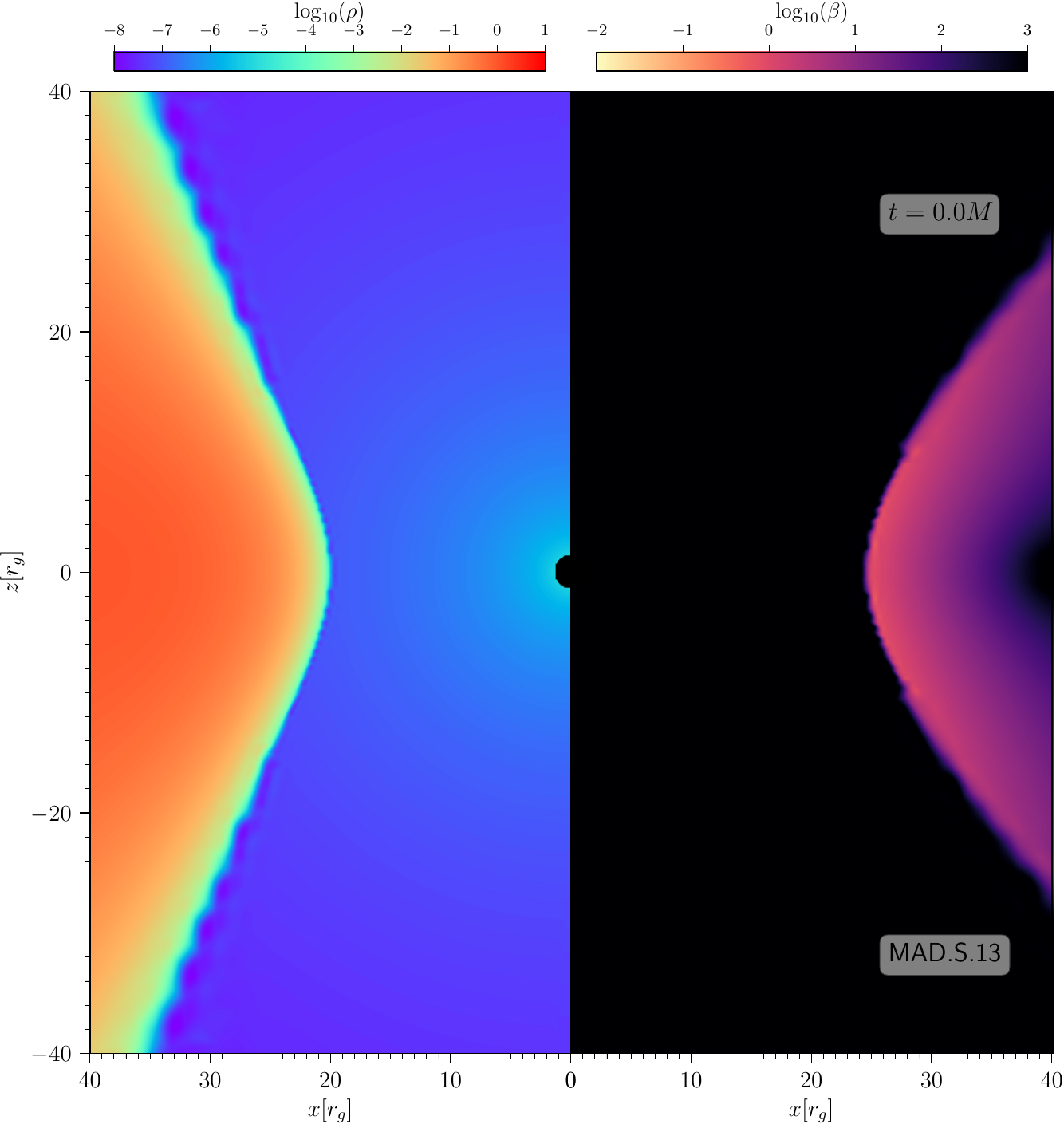}
\includegraphics[width=0.49\textwidth]{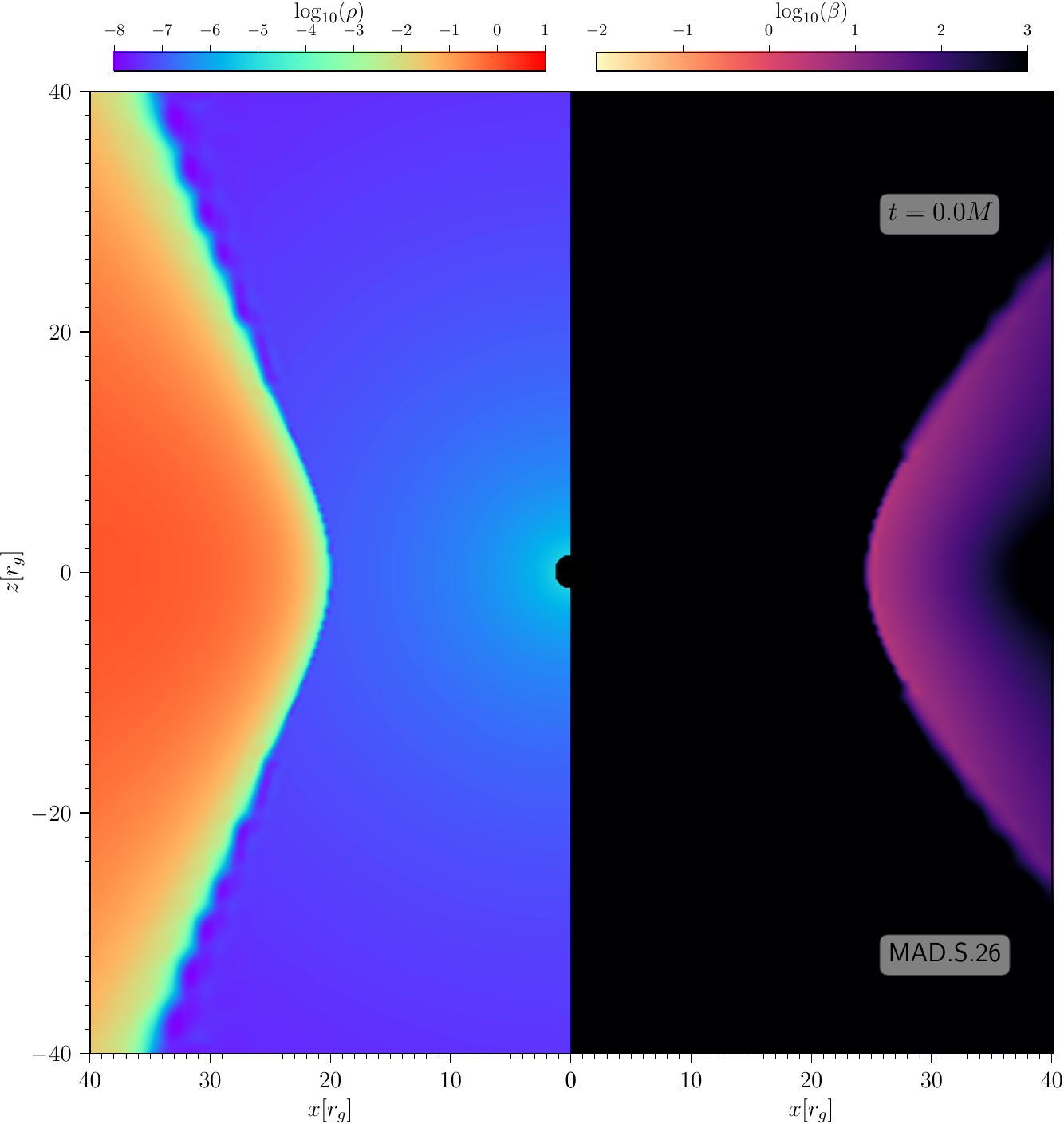}
\includegraphics[width=0.49\textwidth]{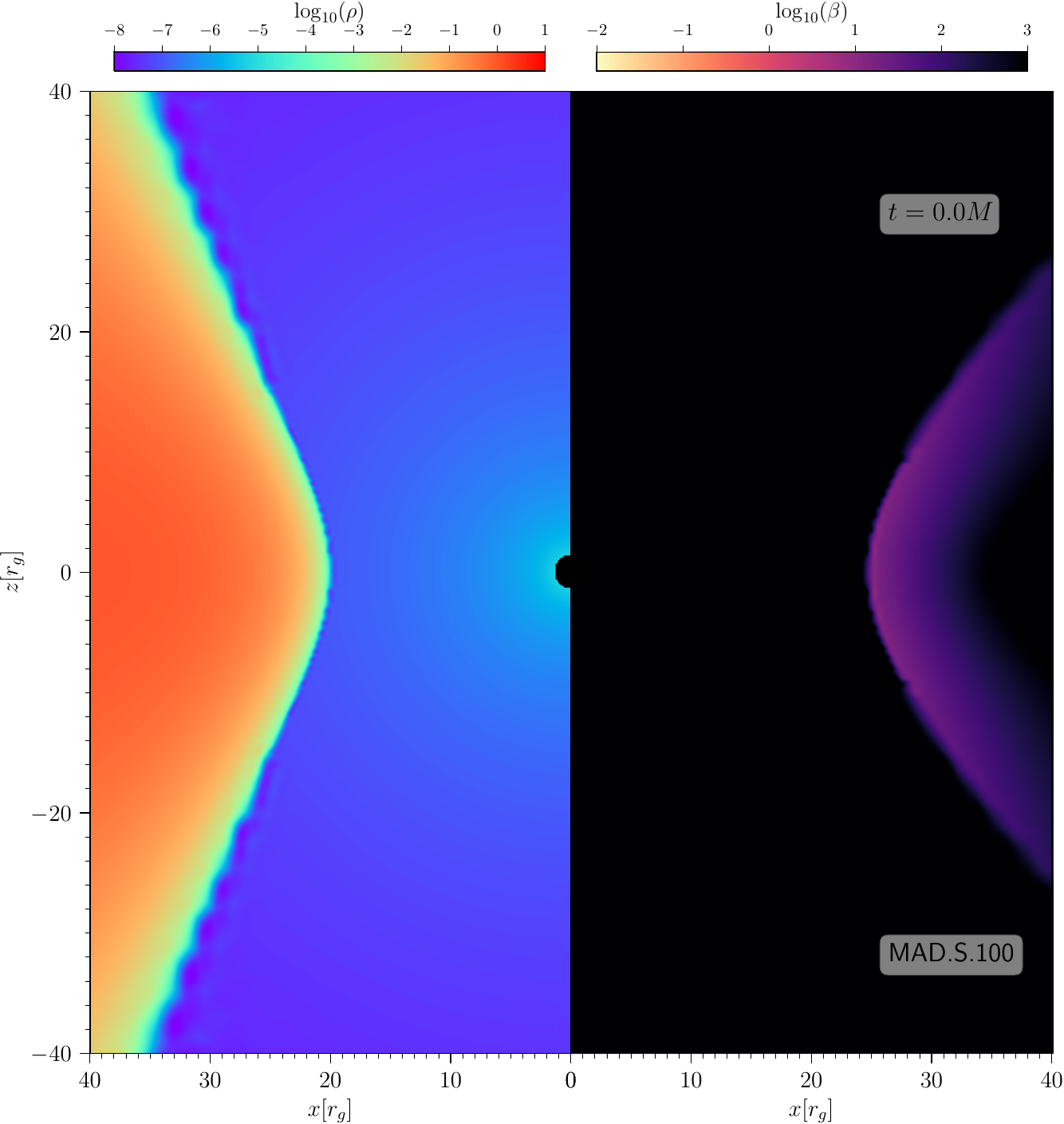}
\caption{A 2D slice of the simulation at initial time, showing rest-mass density  
and plasma-$\beta$. Upper Left panel: model MAD.S.7, Upper Right panel: the extreme model MAD.S.13,
Lower Left panel: model MAD.S.26 and Lower Right panel: model MAD.S.100.}
\label{init}
\end{figure}

The initial snapshot from the all models is seen in Fig.~\ref{init}
where the rest mass density profile and the plasma-$\beta=2p/B^2$ are 
 plotted in logarithmic scale. 

To analyze the main properties of the accretion dynamics we define the rest mass-accretion rate and the magnetic flux across the horizon. The former is measured as:
\begin{align}
 \dot{M}:=\int_0^{2\pi}\int^\pi_0\rho u^r\sqrt{-g}\,d\theta d\phi\,,
\label{mdot} \end{align}
where $\rho$ is the density, $u^r$ is the radial component of the four-velocity and $\sqrt{-g}$ is the determinant of the spacetime metric. Its behaviour is reported as a function of time in the upper part of the left panel of  Figs.~\ref{mdot-p}.  The magnetic flux accreted across the event horizon is defined as: 
\begin{align}
  \Phi_{\rm BH}:=\frac{1}{2}\int_0^{2\pi}\int^\pi_0 |B^r|\sqrt{-g}d\theta d\phi\,,
\label{phiBH} \end{align}
while the ``normalized'' magnetic flux is defined as 
$\phi_{\rm BH}:=  \Phi_{\rm BH}/ \sqrt{\dot{M}}$. 
In Fig.~\ref{initB} we integrate the magnetic flux across $\theta$ and $\phi$ and plot is as a function of r
(as in Eq.~\ref{phiBH}, but for different r).
The most extreme model has much more magnetic flux and as accretion begins most of this flux comes 
to the event horizon of the black hole. 
In the middle (and lower) panels of the left part 
of  Fig.~\ref{mdot-p} the magnetic flux (and 
normalized magnetic flux respectively) is shown 
for all of the models considered. The limiting 
normalized magnetic flux quoted by  \citet{Tchekhovskoy2011} 
is $\phi_{\rm BH}=\phi_{\rm max}\approx 50$. 
In  our simulations using Heaviside-Lorentz units (as opposed to Gaussian  units) this value should be divided by a $\sqrt{4\pi}$, thus $\phi_{\rm  BH}=\phi_{\rm max}\approx 50/\sqrt{4\pi}\approx 14$.

All models considered in this study produce outflows that give off energy at infinity. This can be measured through the energy flux that passes through a 2-sphere placed at $50 \, r_g$
\begin{align}
  P_{\rm jet}:=\int_0^{2\pi}\int^\pi_0 (-T^r_t -\rho u^r)\sqrt{-g}d\theta
  d\phi\,,
\label{pjet} \end{align}
where the integrand in \eqref{pjet} is set to zero if everywhere on the integrating surface $\sigma \leq 1$. The efficiency of the jet is then defined as
\begin{align}
  \eta :=P_{\rm jet}/\dot{M} c^2 \,,
\label{eff} \end{align}
and can become larger than unity (c=1 in code units).

In Fig.~\ref{spin-vs}, we compare the  accumulated magnetic flux at the black hole and 
the efficiency of our super MAD models with values reported in previous studies.
 While previous studies suggested a saturation limit for $\phi_{\rm BH}$, our simulations reveal significantly higher values, particularly in the extreme magnetically arrested disk configurations. Unlike prior works, where the averaging window was 50,000–100,000 $t_{\rm g}$ \citep{Narayan2022}, our averaging is taken from 3,000–10,000 $t_{\rm g}$, during which we observe persistently elevated flux levels. This raises the critical question: Are these extreme flux states merely transient, or do they represent a quasi-steady magnetically arrested disk configuration that can persist indefinitely?

\section{Flux eruption events in super-MAD accretion}

When accretion delivers a sufficient amount of magnetic flux to the black hole, the system reaches a magnetically arrested state. At this point, the normalized magnetic flux on the event horizon ($\phi_{\rm BH} = \Phi_{\rm BH}/\sqrt{\dot{m}} \approx 50$) saturates, and additional magnetic flux can no longer be transported inward by the accreting material \citep{Tchekhovskoy2011}.
Early numerical studies by \citet{Igumenshchev2008} showed that accretion
can continue despite the presence of non-axisymmetric instabilities, which are 
often attributed to interchange processes such as the Kruskal–Schwarzschild or 
magnetized Rayleigh–Taylor modes.
These instabilities facilitate the continued inflow of matter via narrow streams 
that remain connected to the black hole, while episodic eruptions of magnetic 
flux periodically disrupt the local equilibrium on timescales of approximately 
1,000 dynamical times ($t_{\rm g}=r_{\rm g}/c$, where $r_{\rm g}$ is the gravitational radius), such eruptions are not altered in the presence of resistivity \citep{Nathanail2025}.
These well-known flux eruption events are crucial for understanding magnetized accretion and potential flaring activity \citep{Ripperda2022}, and are also essential for interpreting and modeling flares from Sgr A* \citep{Antonopoulou2025}.

Instead of the standard periodic eruptions occurring approximately every 1,000 $t_{\rm g}$, as seen in previous studies, our simulations exhibit large-scale magnetic flux reorganizations spanning wide azimuthal ranges. These events disrupt the disk more globally, suggesting a qualitatively different mechanism of flux accumulation and release compared to conventional magnetically arrested disk dynamics.

\begin{figure}[h]
\centering
\includegraphics[width=0.49\textwidth]{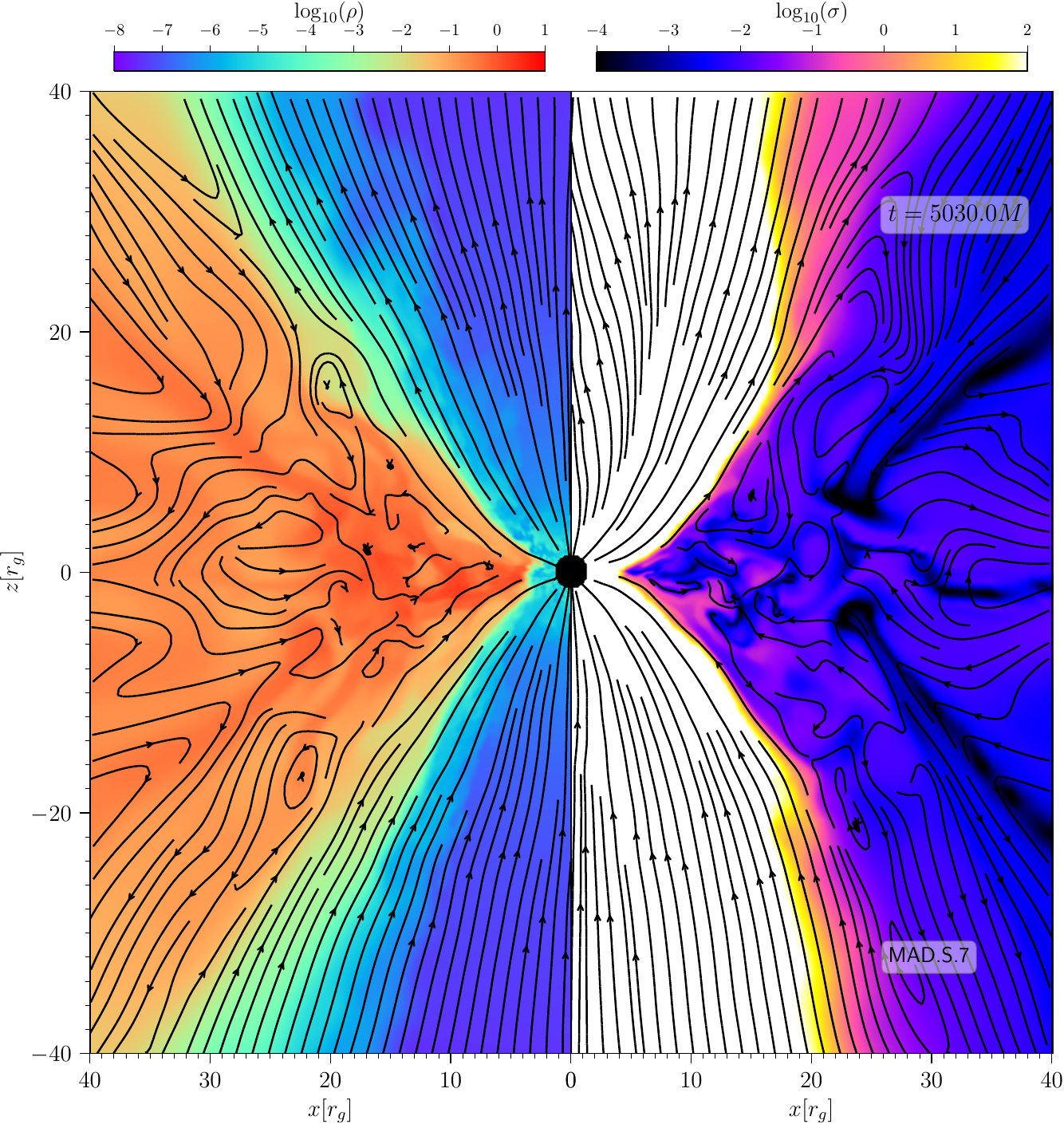}
\includegraphics[width=0.49\textwidth]{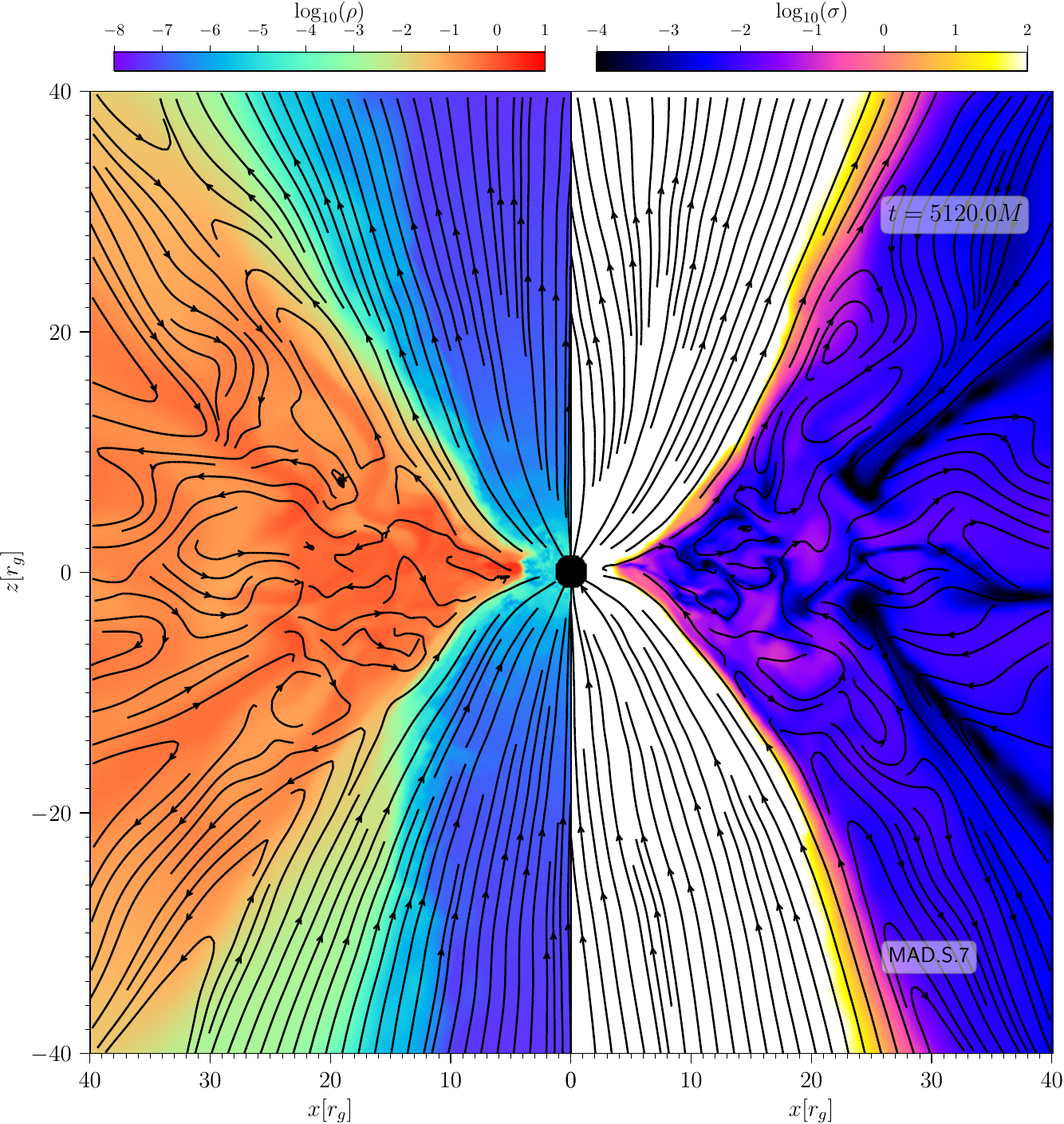}
\includegraphics[width=0.49\textwidth]{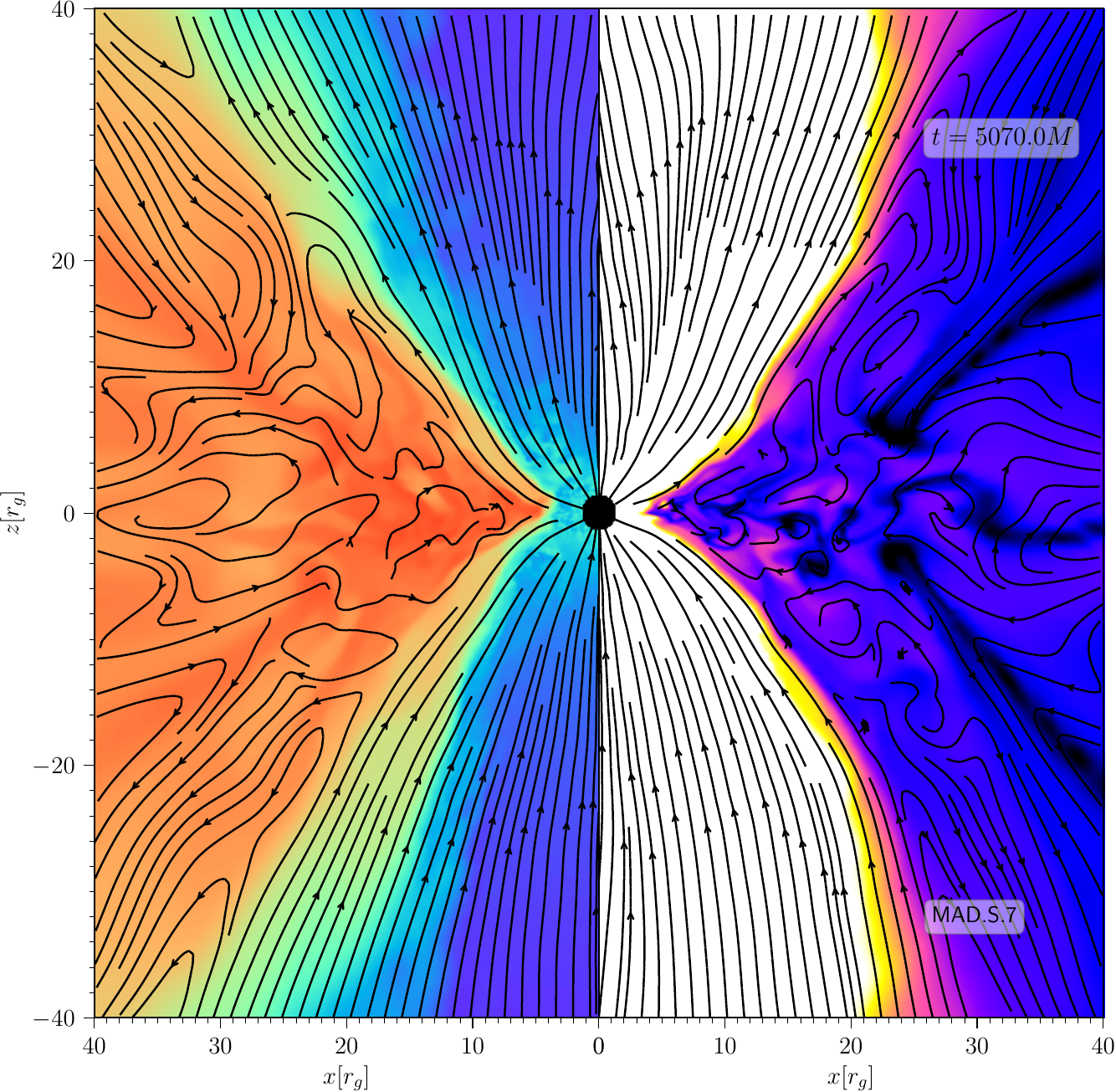}
\includegraphics[width=0.49\textwidth]{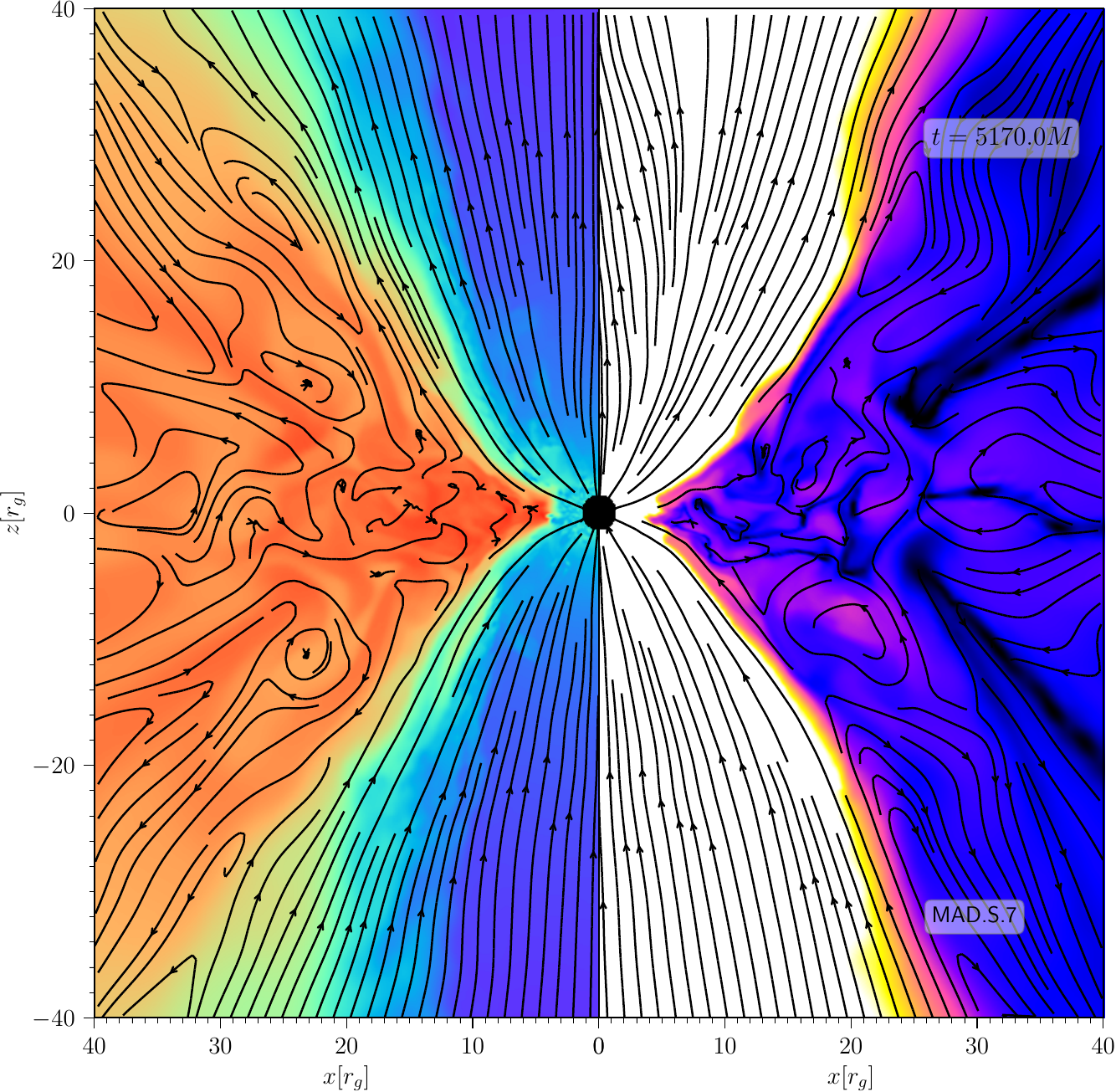}
\caption{ 2D slices of the simulation at constant azimuthal $\phi=180^o$ (left panel) and 
$\phi=0^o$ (right panel) for the extreme model MAD.S.7 from 5,000 - 5,170 $t_{\rm g}$. Left panel depicts the rest mass density whereas the right panel the magnetization $\sigma$.}
\label{eruption}
\end{figure}

In model MAD.S.13 exhibits much higher variability. It undergoes multiple large-scale eruption events that temporarily suppress accretion, although for significantly shorter durations ($\sim$100–200 $t_{\rm g}$; see also the accompanying animation for  model \href{https://zenodo.org/records/15745329?token=eyJhbGciOiJIUzUxMiJ9.eyJpZCI6IjI2YzQ5NDg3LTRmM2EtNGQ1NC04NGVmLTkxMzc2MzM5YWMyNCIsImRhdGEiOnt9LCJyYW5kb20iOiI4YzlkZjJlMmY4YTAzMWUwY2YzYjMzZWFlMzY3NGM4NSJ9.fNqJn0kodbH1q7m6FpHITICHW4mOsZkPcLNcTPiIydaH76QNPMjXKchFsWEfhmP6AAL5YaAYXJERbye2muRl-w}{MAD.s.7} and \href{https://zenodo.org/records/15745251?token=eyJhbGciOiJIUzUxMiJ9.eyJpZCI6Ijg4ZTc2ZWRhLWU2MzItNGU3Ny1hZTcwLTViYmI4ODBjZWNjNSIsImRhdGEiOnt9LCJyYW5kb20iOiJhMjZlMGY5ZTI0MDQyOGNjZTE1NzcxMjQxMWNhZTBlZCJ9.K5D5dAZdwx99JE0SKBVAsbq4p0rZ-KZFYw--D4to9X88iLvcSkpyyUBU5p0B7z30n31dRUIkXX4Sa1qxOMG5TQ}{MAD.s.13}). The mass accretion rate in this model can fluctuate by up to three orders of magnitude. Small-scale variations are also seen in the absolute magnetic flux threading the horizon, suggesting that accretion attempts to deliver more magnetic flux. 
It appears that a strong initial eruption is followed by a period of recovery, during which the disk succeeds in sending small streams of accreting material to the black hole. These streams exhibit low azimuthal mode numbers ($m=2$ or $m=4$), but are much weaker and narrower than the standard MAD streams commonly seen in prior simulations.

By contrast, in model MAD.S.7, we observe a major flux eruption event that lasts for approximately 3,000 $t_{\rm g}$ and spans the entire azimuthal extent, resulting in a  halt of accretion, the accretion disk is unable to reach the black hole during this time. Fig. \ref{eruption} 
follows this event for 170 $t_{\rm g}$ and is a continuation of Fig. \ref{vert}

Model MAD.S.26 is the least variable among the extreme MAD configurations. The magnetic flux at the horizon remains relatively stable over time, and the accretion rate shows milder fluctuations compared to MAD.S.13 and MAD.S.7. Its eruption dynamics lie between those of a standard MAD and the more extreme models discussed here. It resembles a system in which a modest flux eruption has occurred, but accretion continues via narrow, non-axisymmetric instabilities such as exchange or detachment modes. Notably, the morphology of these accretion streams differs from those in standard MADs, as they remain more confined and localized near the black hole.

\begin{figure}[h]
\centering
\includegraphics[width=0.59\textwidth]{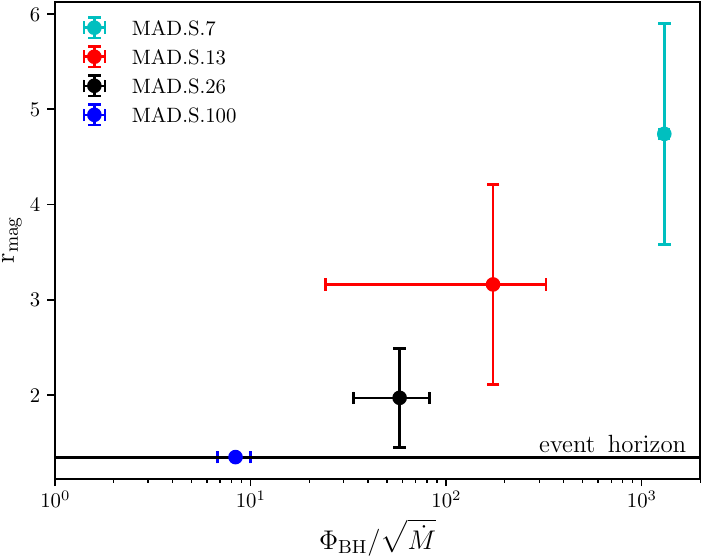}
\caption{ For each model,  we show the mean magnetospheric radius (vertical axis), defined from the equatorial 
$\sigma=1$ contour, as a function of the normalized magnetic flux at the horizon (horizontal axis). Error bars denote the 
$1-\sigma$ azimuthal variation over a time window of $1000 \,\,\rm {t_g}$. The black horizontal line marks the event horizon. The standard MAD model (MAD.s.100) remains pinned to the horizon, while MAD.s.23 shows intermittent outward excursions. The more extreme models MAD.s.13 and MAD.s.7 sustain significantly larger standoff radii (
$\sim 3\,\, r_{\rm g}$ and $\sim 5 \,\, r_{\rm g}$, respectively), exhibiting very large azimuthal extent, indicative of a persistent global flux eruption state.}
\label{rflux}
\end{figure}

To quantify the disk standoff in each model, we define the \textit{equatorial magnetospheric radius}, $r_{\rm mag}$, as the radial position where the magnetization goes to unity, $\sigma(r,\phi,\theta=\pi/2)=1$. 
For each time slice, we extract the equatorial plane, locate the $\sigma=1$ contour, and measure the radial distance $r_{\rm mag}(\phi,t)$ at each azimuth $\phi$. 
We then compute the azimuthal mean $\langle r_{\rm mag}(t)\rangle_{\phi}$ and its $1\sigma$ standard deviation. Finally, these values are averaged over a $1,000\,t_{\rm g}$ window (from $4,000$--$5,000\,t_{\rm g}$) to provide $\langle r_{\rm mag}\rangle$ for each model, paired with the respective time-averaged normalized magnetic flux at the horizon, $\langle \Phi_{\rm BH}/\sqrt{\dot{M}}\rangle$. These results are shown in Fig.~\ref{rflux}.

In the standard MAD (MAD.s.100), $r_{\rm mag}$ remains at the horizon, consistent with previous studies. By contrast, models with higher accumulated magnetic flux exhibit progressively larger magnetospheric radii: $\sim 2\,r_{\rm g}$ in MAD.s.23, $\sim 3\,r_{\rm g}$ in MAD.s.13, and $\sim 5\,r_{\rm g}$ in MAD.s.7. This systematic trend demonstrates that the sustained clearing of the inner accretion flow is physically controlled by horizon magnetic flux.
As shown in Fig. \ref{rflux}, the mean equatorial magnetospheric radius increases systematically with $\Phi_{\rm BH}/\sqrt{\dot{M}}$, reaching $\sim 5\,r_{\rm g}$ in the most extreme case. 

me magnetic flux is robust to variation in the floor density.

\begin{figure}[h]
\centering
\includegraphics[width=0.76\textwidth]{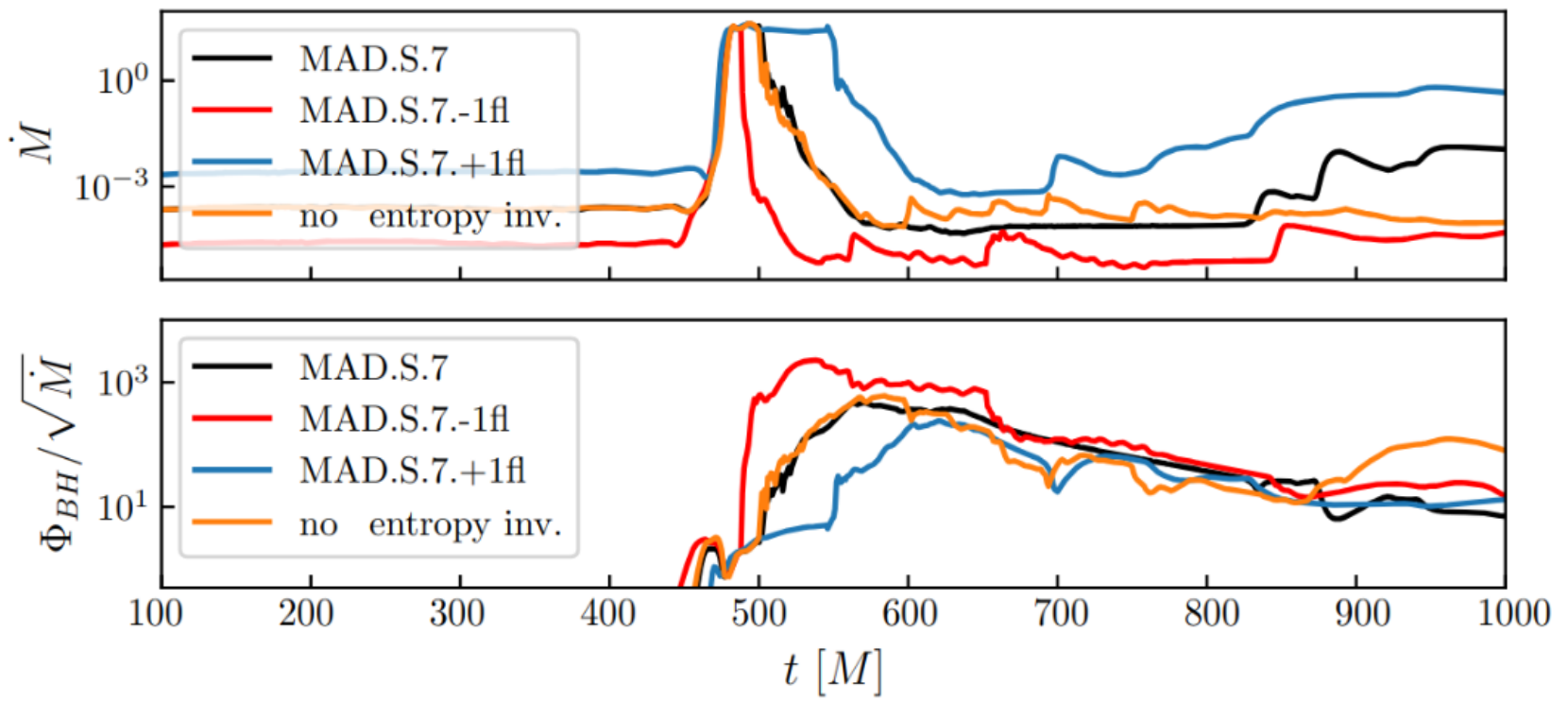}
\caption{Upper panel: evolution of the mass accretion measured across the black hole event horizon. Lower panel: evolution of the normalized magnetic flux accreted onto the black hole.  It is for the extreme model MAD.s.7, for one order of magnitude 
lower floor density (MAD.S.7.-1fl), similarly for larger floor density (MAD.S.7.+1fl), and the same MAD.s.7.}
\label{mdot_r}
\end{figure}

We further assessed the robustness of these results with respect to the primitive-variable inversion strategy. Using the same reduced resolution, we carried out two additional simulations based on MAD.s.7: one employing BHAC’s default treatment, which switches to solving the entropy equation below a critical plasma-beta threshold \citep{Porth2017}, and another in which the entropy inversion was disabled entirely, relying exclusively on the energy-based inversion. As shown in Fig.~\ref{mdot_r}, both runs display comparable evolution, with the normalized magnetic flux in the energy-only inversion case reaching slightly higher values. This confirms that the emergence of a highly magnetized funnel, significantly exceeding the magnetization levels of standard magnetically arrested disk simulations, is not sensitive to the inversion scheme and represents a robust physical outcome of the system’s evolution.

\begin{figure}[h]
\centering
\includegraphics[width=0.32\textwidth]{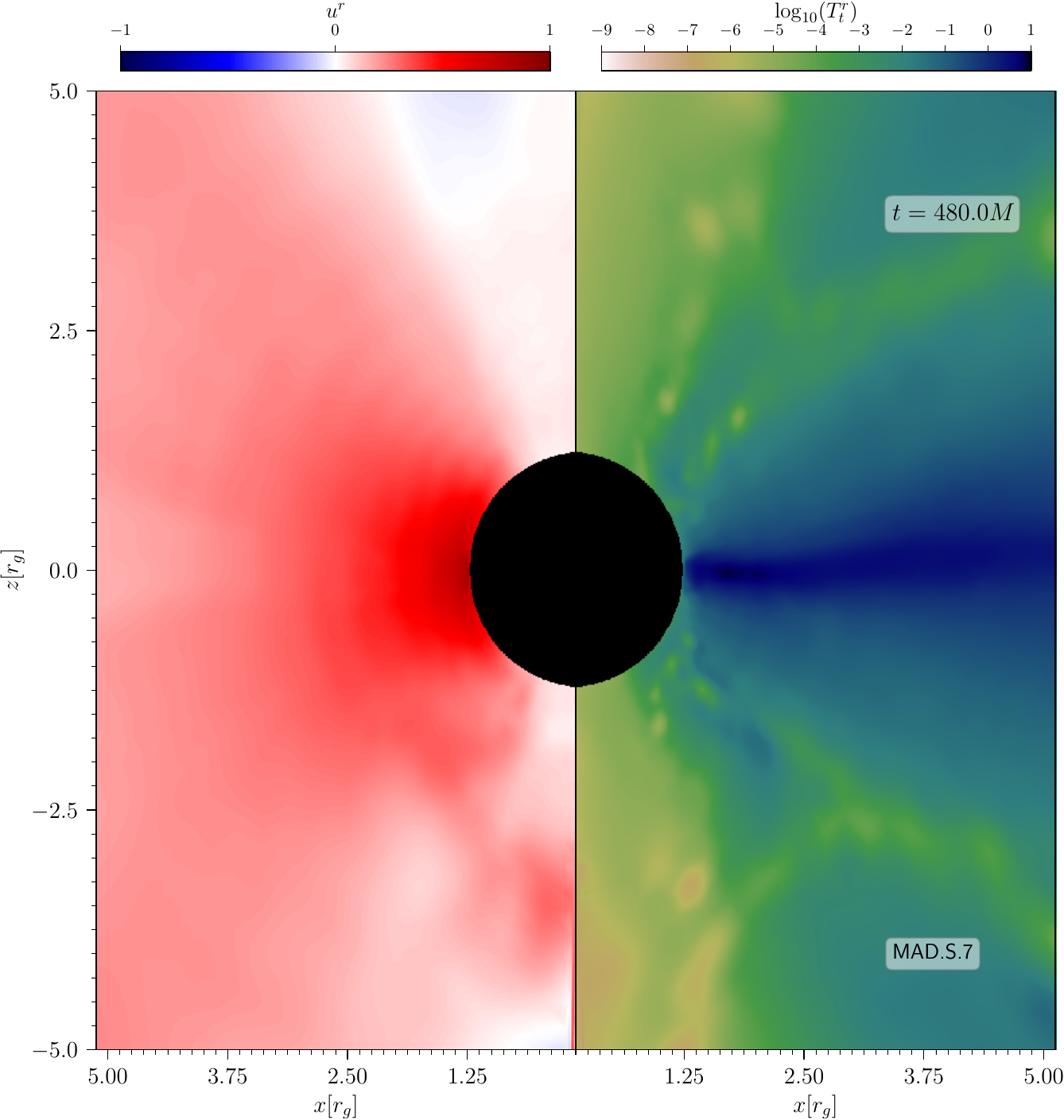}
\includegraphics[width=0.32\textwidth]{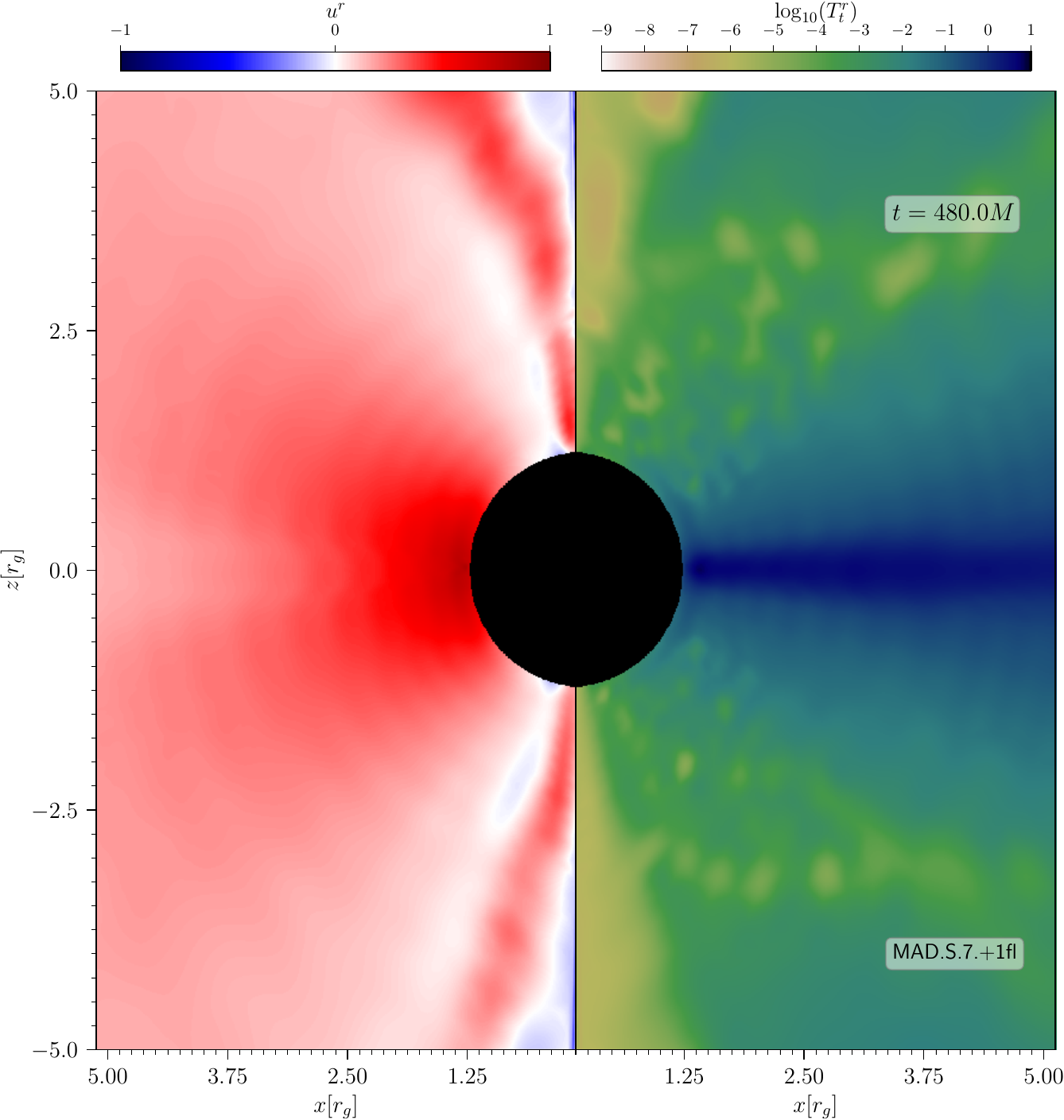}
\includegraphics[width=0.32\textwidth]{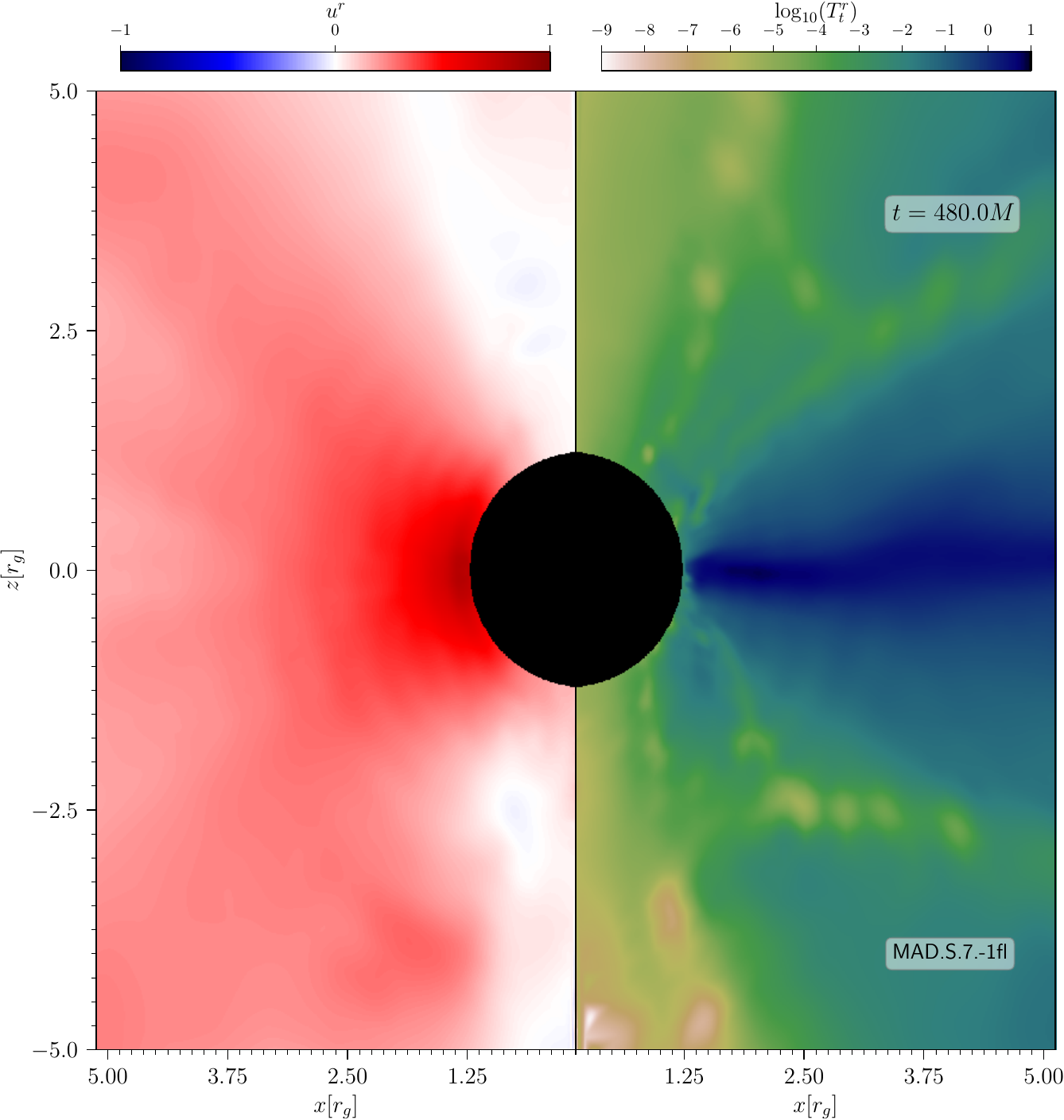}
\includegraphics[width=0.32\textwidth]{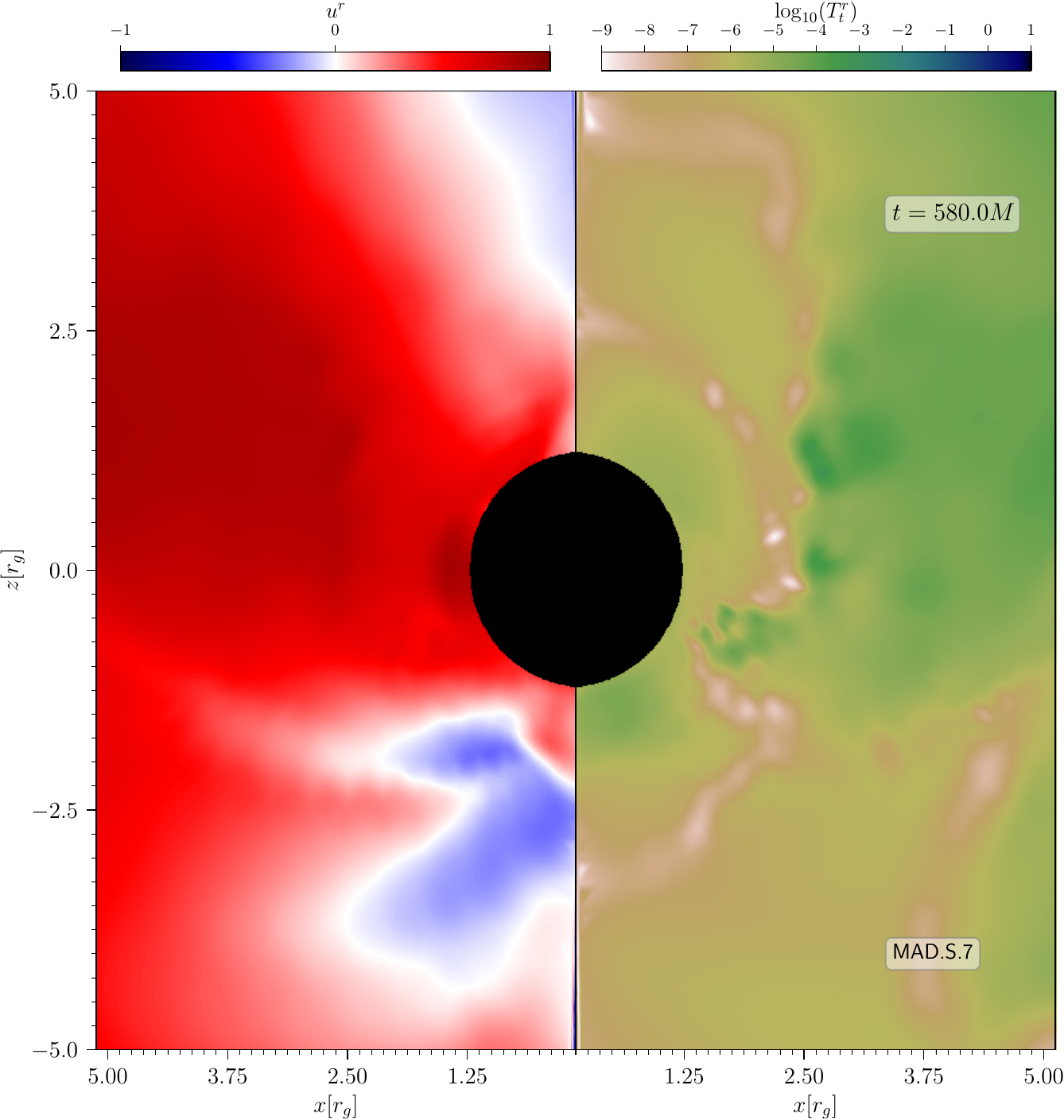}
\includegraphics[width=0.32\textwidth]{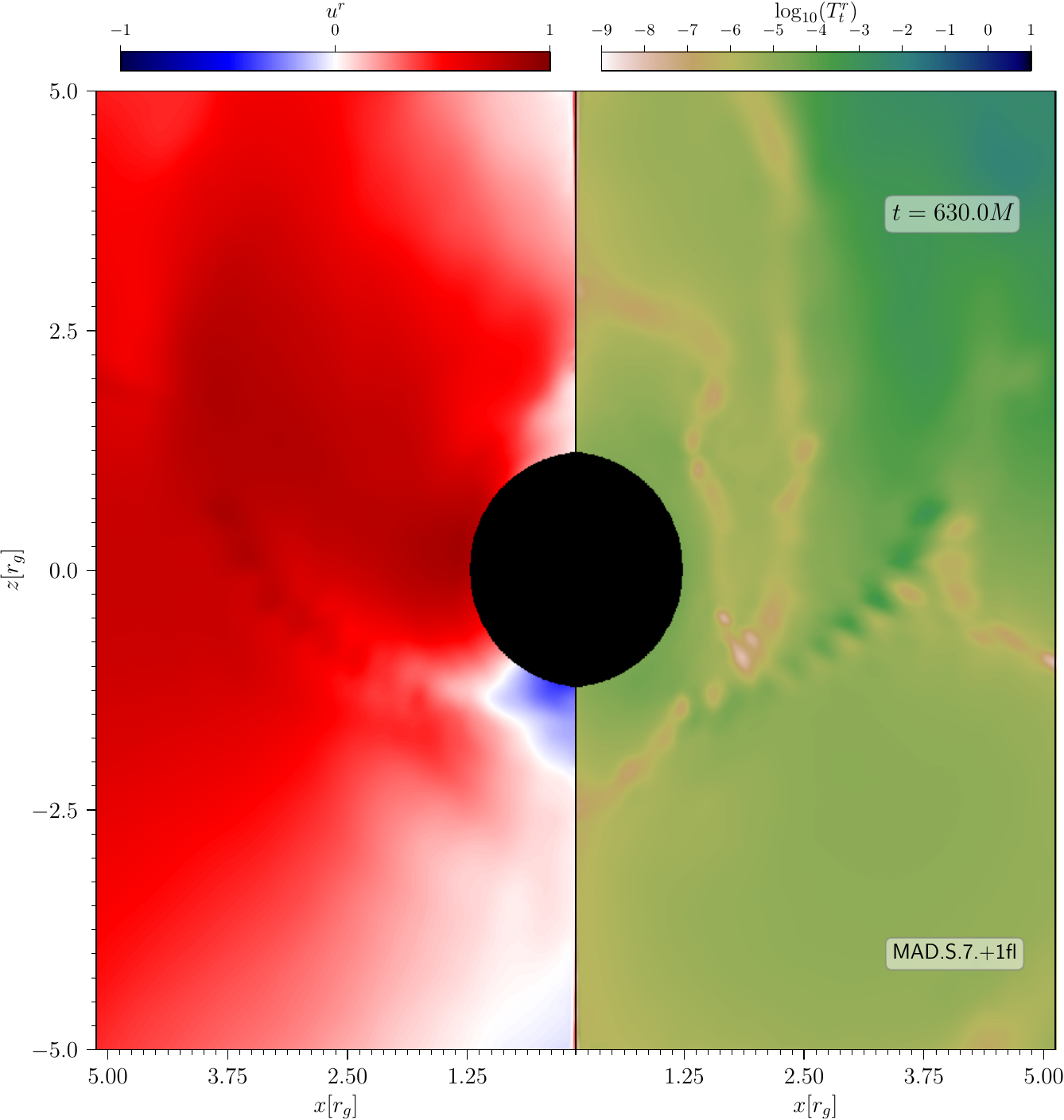}
\includegraphics[width=0.32\textwidth]{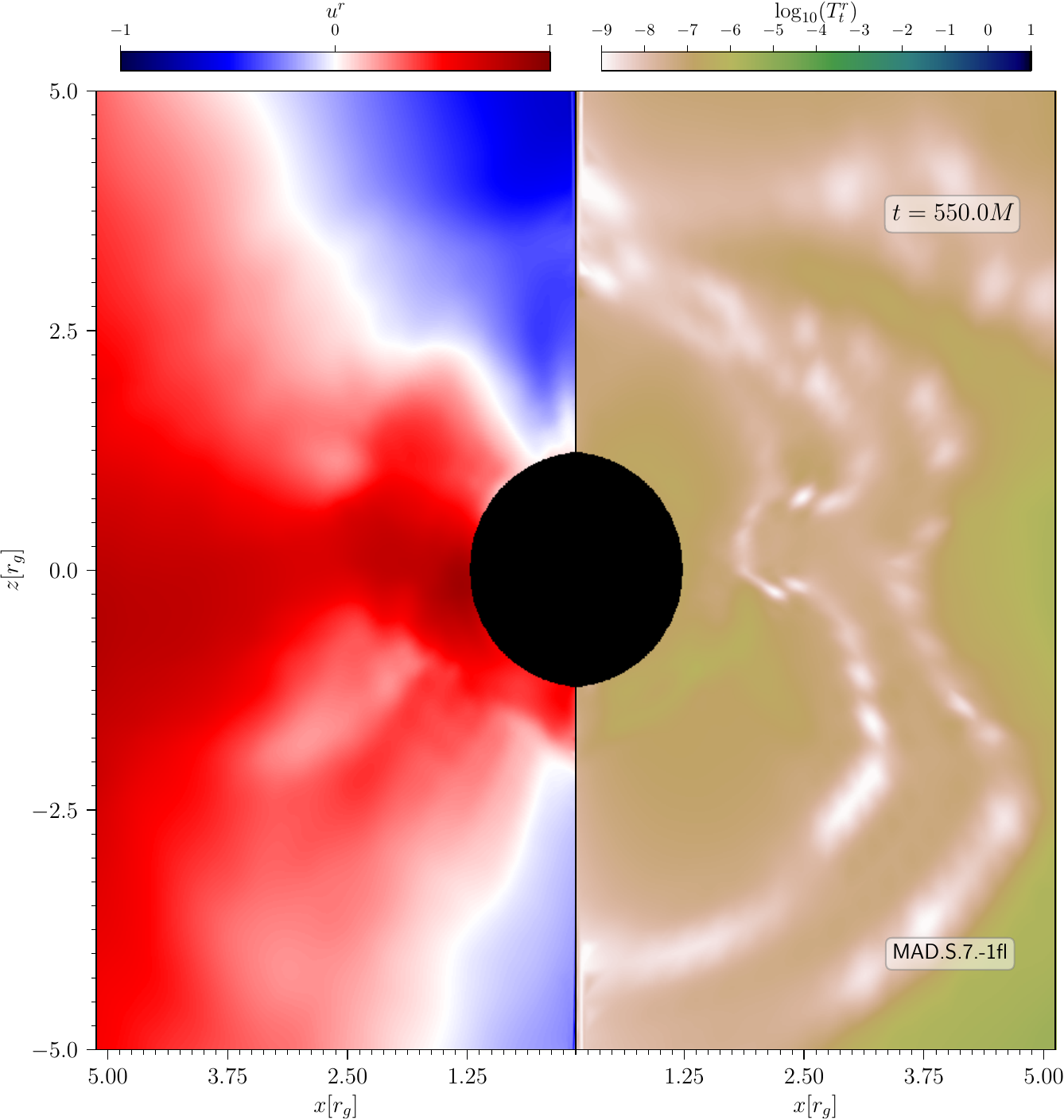}
\caption{Snapshots of the radial kinetic energy flux  (right half-panels) and radial velocity (left half-panels) are shown for three simulations with different floor density levels. Each column corresponds to a different model:  standard (Left column), 10× higher (Center), and 10× lower (Right). The upper row indicates an accretion state at the same time for each model, whereas the lower row shows the time of peak magnetic flux eruption for that specific model, allowing consistent comparison despite slight timing differences. }
\label{Trt}
\end{figure}

To further evaluate the influence of the numerical floor on accretion behavior near the black hole, we analyzed the kinetic part of  the radial energy flux component of the stress-energy tensor, $T^r_t$, which captures the flow of kinetic energy, along with the radial component of the fluid velocity, $u^r$. We examined two representative snapshots from each simulation, one during active accretion and another one at the peak of a strong flux eruption, in order to assess the dynamical impact of the floor. 
For each model, the timing of the peak eruption differs slightly, therefore, we select the respective peak moment for each simulation to enable a meaningful comparison.
Fig.~\ref{Trt} shows the  results for the three lower resolution models: one with the default floor density, one with a floor value an order of magnitude higher, and one an order of magnitude lower
(MAD.s.7, MAD.s.+1fl, MAD.s.7.-1fl, respectively).
The upper row of Fig.~\ref{Trt} corresponds to a phase of active accretion. Across all floor prescriptions, the velocity profiles and the
associated radial kinetic energy fluxes are broadly similar, indicating
consistent inflow behavior during this state. The lower row captures
the moment of peak magnetic flux eruption for each model, note that the 
timing of this event differs slightly across simulations due to their
individual evolution.

In the high-floor model, polar outflows are suppressed, and
nearly all material is funneled directly into the black hole. As the 
floor density is reduced, polar outflows begin to emerge, becoming more 
pronounced in the lowest-floor case. In the radial energy flux maps, 
small localized anomalies appear near the horizon, originating from 
failed cells that quickly equilibrate with the surrounding flow. The 
magnitude of the kinetic energy flux in all cases is negligible, compared 
with the upper row accretion state. However, we need to mention that it 
is sensitive to the floor value, 
with lower floor densities yielding more physically realistic outflow 
structures and less kinetic energy radial flux inflow.  
Overall, we find that the regions 
where the floor is activated contribute negligibly to the total radial 
kinetic energy flux.

The upper row corresponds to active accretion state and for all models 
we have a similar velocity profile and similarly a kinetic energy flux 
profile. In the lower row the peak of the flux eruption event is 
depicted for each model, note that in each model this happens at 
a slightly different time. In the high floor model, any outflow seems to be suppressed as all material goes to the black hole, as we go to less and less floor density, a polar outflow is produced. In the radial energy 
flux profile there are regions that come from a failed cell which 
however quickly cacthes the surounding material velocity , and the value of the kinetic energy flux depends a lot on the floor density. Clearly going to smaller and smaller floor is expected to go closer to physical realtity. Thus the outflow in the lower floor model and the low inward 
kinetic energy flux we believe are fine.
Across all cases, we find that the regions where the floor is activated contribute negligibly to the total radial kinetic energy flux.

\section{The quality factor of the MRI }
In order to be sure that we have enough resolution to resolve the 
magneto-rotational instability MRI we 
employ the  magneto-rotational  MRI quality 
factor $Q_\theta$ which we define below, and details on its calculation for all simulated models presented 
in Table~\ref{table:models}. We evaluate the so-called ``quality factor'' $Q_\theta$,
in terms of the ratio between the grid spacing in a given direction
$\Delta x_{\theta}$, (eg the $\theta$-direction) and the wavelength of
the fastest growing MRI mode in that direction (ie $\lambda_{\theta}$).
Both quantities are evaluated in the tetrad basis of the fluid
frame $e_{\mu}^{(\hat{\alpha})}$ (see \citealt{Takahashi:2008,
  Siegel2013, Porth2019}, for details)
\begin{align}
  Q_{\theta}:=\frac{\lambda_{\theta}}{\Delta x_{\theta}}\,,
\label{Qth}
\end{align}
where
\begin{equation}
\lambda_{\theta}:=\frac{2\pi}{\sqrt{(\rho h +b^2)}\Omega}
b^{\mu}e_{\mu}^{(\theta)}\,,
\end{equation}
$\Omega:=u^{\phi}/u^t$ is the angular velocity of the fluid and the
corresponding grid resolution is $\Delta x_{\theta}:=\Delta
x^{\mu}e_{\mu}^{(\theta)}$. The average of $Q_{\theta}$ is done in space and  time, specifically in a time window of 200 M and spatially in the region of interest at angles $60^o < \theta < 120^o$ and $r<40\, r_g$,
inside the heart of the disk.

\section{Discussion}\label{sec12}

Our simulations  indicate a spectrum of magnetically arrested behaviors rather than a single transient outcome. In two of the models, the accumulation of magnetic flux leads to a long-lived displacement of the inner accretion flow, with the magnetospheric radius remaining at $\sim 3,r_{\rm g}$ for more than $10{,}000,t_{\rm g}$. In the most extreme case, the system undergoes a stronger but shorter-lived episode in which the magnetosphere expands to $\sim 5,r_{\rm g}$, temporarily halting accretion over the full azimuthal extent. Whether configurations exist that can sustain such large magnetospheric radii for longer timescales, or push the magnetosphere even farther outward, cannot be determined from the present study and remains an open question for future work.
Such extreme, magnetically dominated states may nonetheless be realized in nature during highly dynamical events. In the context of tidal disruption events, for example, the presence of a strong poloidal magnetic field has been shown to be essential for reproducing the jet properties observed in Swift J1644+57 \citep{Tchekhovskoy2014, Curd2019}.
However, it remains uncertain whether such poloidal 
fields are generated solely by the accretion flow’s turbulent dynamo or whether they 
require pre-existing magnetic flux, possibly carried by the infalling disrupted star  \citep{Guillochon2017b, Bonnerot2017}. 
Interestingly, recent high-
resolution simulations suggest that even purely toroidal magnetic fields, which could be 
present in the accretion flow during a tidal disruption event, are capable of leading to the formation of a 
magnetically arrested disk  state \citep{Liska2018}. This underscores the 
possibility that strong field configurations may form through accretion dynamics 
themselves, independent of pre-existing field structures. A similar situation may arise in 
the case of gamma-ray bursts, where the enormous magnetic field within the interior 
of a star is accreted onto the black hole, leading to the formation of such transient 
events.

\bibliography{samlpe701}{}
\bibliographystyle{aasjournalv7}



\end{document}